\documentclass[transmag]{IEEEtran}
\usepackage{amsfonts}

\usepackage{epsfig}
\usepackage{graphicx}
\usepackage{psfig}
\usepackage{subfigure}
\usepackage{epsf}
\usepackage{epstopdf}
\usepackage{cite}
\usepackage{amsmath,amssymb,amsfonts}
\usepackage{algorithmic}
\usepackage{graphicx}
\usepackage{textcomp}
\usepackage{booktabs}
\usepackage{fancyhdr}

\usepackage{algorithm}
\usepackage{algorithmic}
\usepackage{epstopdf,cite,color}
\usepackage{amsthm}
\usepackage{amssymb}
\newcommand{\bc}{\begin{center}}
	\newcommand{\ec}{\end{center}}
\newcommand{\be}{\begin{equation}}
\newcommand{\ee}{\end{equation}}
\newcommand{\bea}{\begin{eqnarray}}
\newcommand{\eea}{\end{eqnarray}}

\usepackage[letterpaper, left=0.625in, right=0.625in, bottom=1in, top=0.70in]{geometry}

\renewcommand\IEEEkeywordsname{Index Terms}

\def\BibTeX{{\rm B\kern-.05em{\sc i\kern-.025em b}\kern-.08em T\kern-.1667em\lower.7ex\hbox{E}\kern-.125emX}}
\begin{document}
	\title{When Machine Learning Meets Spectrum Sharing Security: Methodologies and  Challenges}
	
\author{Qun~Wang \IEEEmembership{Student Member, IEEE}, , Haijian Sun \IEEEmembership{Member, IEEE},\\
Rose~Qingyang~Hu \IEEEmembership{Fellow, IEEE}, Arupjyoti Bhuyan \IEEEmembership{Senior Member, IEEE}

\thanks{Manuscript received November 2, 2021. This project is sponsored by Idaho National Laboratory.}
\thanks{Q. Wang and R. Q. Hu are with the Department
		of Electrical and Computer Engineering, Utah State University, Logan, UT, USA. E-mails: claud.wang,  rose.hu@usu.edu. 
}
\thanks{	H. Sun is with the Department of Computer Science, University of Wisconsin-Whitewater, WI, USA.
		Email: sunh@uww.edu.}
\thanks{A. Bhuyan is with the Wireless Security Institute, Idaho National Laboratory. Email: arupjyoti.bhuyan@inl.gov.}}
	
		


\IEEEtitleabstractindextext{\begin{abstract}

The exponential growth of internet connected systems has generated numerous challenges, such as spectrum shortage issues, which require efficient spectrum sharing (SS) solutions. {Complicated and dynamic SS systems can be exposed to different potential security and privacy issues, requiring protection mechanisms to be adaptive, reliable, and scalable.} Machine learning (ML) based methods have frequently been proposed to address those issues.
In this article, we provide a comprehensive survey of the recent development of ML based SS methods, the most critical security issues, and corresponding defense mechanisms. 
In particular, we elaborate the state-of-the-art methodologies for improving the performance of SS communication systems for various vital aspects, including ML based cognitive radio networks (CRNs), ML based database assisted SS networks, ML based LTE-U networks, ML based ambient backscatter networks, and other ML based SS solutions. We also present security issues from the physical layer and corresponding defending strategies based on ML algorithms, including Primary User Emulation (PUE) attacks, Spectrum Sensing Data Falsification (SSDF) attacks, jamming attacks, eavesdropping attacks, and privacy issues. Finally, extensive discussions on open challenges for ML based SS are also given. This comprehensive review is intended to provide the foundation for and facilitate future studies on exploring the potential of emerging ML for coping with increasingly complex SS and their security problems.
\end{abstract}

}

\maketitle

\section{Introduction}
\IEEEPARstart{T}{he} explosion of data traffic growth, massive device increment, and commercialization of the 5G wireless communication networks impose great challenges on spectrum efficiency (SE) as well as data security \cite{5ghu}. On the one hand, in order to meet requirements in the 5G communication system, the future wireless solutions should provide a $10-100$ times higher data rate and support a $10-100$ times higher density of connected devices. {On the other hand, the spectrum fragmentation and the fixed allocation policies adopted in traditional wireless networks do not make efficient use of the scarce spectrum resource and limit the possibility of meeting such a high capacity requirement.}
Furthermore, the complicated communication environment brings more risks to the users and systems. Thus security and privacy issues have become the primary concern in 5G networks.

The SS network can help relieve the shortage of spectrum resources.
Different from traditional exclusive frequency allocations, SS by definition involves multiple entities and uses the spectrum in a shared way in order to increase the efficiency of the limited spectrum resources. 
According to \cite{cls1}, SS can fall into two main categories: horizontal sharing and vertical sharing. {In horizontal sharing, it implies that all the networks and users have equal rights to access the spectrum.} Such methods allow the users to co-exist peacefully and efficiently. {Vertical sharing, on the other hand, allows multi-type users to access the spectrum resources with different rights.} Therefore, secondary users (SUs) can use the spectrum without harming the performance of primary users (PUs). By enabling SUs to access the spectrum owned by the PUs, the limited spectrum resource can support more devices.

One of the technical challenges in the vertical SS system is how to guarantee the performance of different types of users while achieving the highest SE. To this end, the spectrum access mechanism, interference control, resource allocation, and fairness all need to be tackled in a dynamic and collaborative way. Since the concept of SS was first introduced, different SS frameworks based on various application scenarios have been developed by researchers. A brief history of SS networks is presented as follows.

\subsection{SS: A Brief History}
The concept of vertical SS was first brought up by decentralized and opportunistic cognitive radio (CR) techniques, where SUs exploited the idle spectrum with sensing ability to transmit their information without causing any harmful interference to the licensed PUs \cite{sshs01}. 

Traditional CR techniques enabled SUs to take advantage of spectrum opportunities by learning/monitoring the environment and adjusting their transmission parameters adaptively. { However, the opportunistic access to the bands without a license in CRN makes it challenging to guarantee the quality of service (QoS) and maintain a low level of interference when multiple service providers co-exist. In particular, with the increase of primary wireless devices and activities in the network, SUs may have very limited access opportunities to the spectrum resource. }
{Therefore, new licensed spectrum access methods are needed to provide more predictable and controllable solutions to meet the high QoS requirement for both SUs and PUs \cite{licesy1}.}

In 2012, Qualcomm and Nokia initially proposed the concept of Authorized Shared Access (ASA), which was further extended to the Licensed Shared Access (LSA) framework by several European institutions, including the European Conference of Postal and Telecommunications Administrations (CEPT), the European Commission (EC), and the Radio Spectrum Policy Group (RSPG) of the European Union (EU). The main objective of LSA is to allow new users to work in already occupied frequency bands while maintaining existing incumbent services on a long-term basis. The LSA framework is currently switched to explore the 3.4-3.8 GHz band from the 2.3-2.4 GHz band, enabling coexistence between incumbents and 5G applications \cite{sshs02}\cite{sshs03}. 

Similar to the LSA framework, the US Federal Communications Commission (FCC) proposed the Citizens Broadband Radio Service (CBRS) in order to open and share the frequency band 3.55-3.7 GHz. It sought to improve spectrum usage by allowing commercial users to share the band with incumbent military radars and satellite earth stations. 
{In this approach, the access and user coordination are controlled by the corresponding Spectrum Access System (SAS).} SAS comprises three types of users with different levels of priority. The first type is Incumbent Users (IUs), which have {the highest spectrum access priority}. The second is the Priority Access License (PAL) users, which can exclusively access the spectrum without the existence of  IUs. The third is General Authorized Access (GAA) users, who have sensing-assisted unlicensed access in the absence of the incumbent and PAL users. {SAS determines the maximum allowable transmission power level and the available frequencies at a given location to be assigned to PAL and GAA users \cite{sshs02}\cite{sshs03}.}

It should be noted that both LSA and SAS systems are defined for usage in a specific frequency band. LSA is mainly based on database-assisted SS, while SAS combines a database with Environmental Sensing Capability (ESC) protections. Thus in SAS,  the radio resources allocation decision is obtained with assistance from the spectrum database and sensing results. The ESC can more effectively protect IUs from harmful interference while guaranteeing their privacy. The database can provide a more stable service for SUs than CRN \cite{sshs033}.

{Much research has been devoted to increasing licensed systems' capacities by the extension of Long Term Evolution (LTE) over the unlicensed spectrum band.}  Several concepts have been proposed, such as LTE in unlicensed bands (LTE-U), License Assisted Access (LAA) in LTE Advanced (LTE-A), LTE Wireless Local Area Network (WLAN) Aggregation (LWA), etc. By allowing LTE users to operate on the unlicensed band without causing any harmful interference to original users such as WiFi devices, coexisting technologies can have a great impact on the spectrum access in the immediate future \cite{sshs02}.

The emergence of Internet of Things (IoT) networks presents new challenges to wireless communication design from both spectrum and energy aspects. Supporting communication with power-limited IoT devices, Ambient Backscatter Communication (AmBC) has attracted extensive attention as a promising technology for SS communications. In AmBC systems, backscatter devices can use surrounding signals from ambient RF sources to communicate with each other. By modulating and reflecting surrounding ambient signals, the backscatter transmitter can transmit data to the receiver without consuming new spectrum resources. The receiver can decode and obtain useful information after receiving the signal. Therefore, the AmBC system does not require a dedicated frequency spectrum, and the number of RF components is minimized at backscatter devices. {Those devices can transmit data with sufficient harvested energy from RF sources \cite{amb01}, which can also improve system energy efficiency significantly.}
\begin{figure}[h]
	\centering
	\includegraphics[width=3.0in]{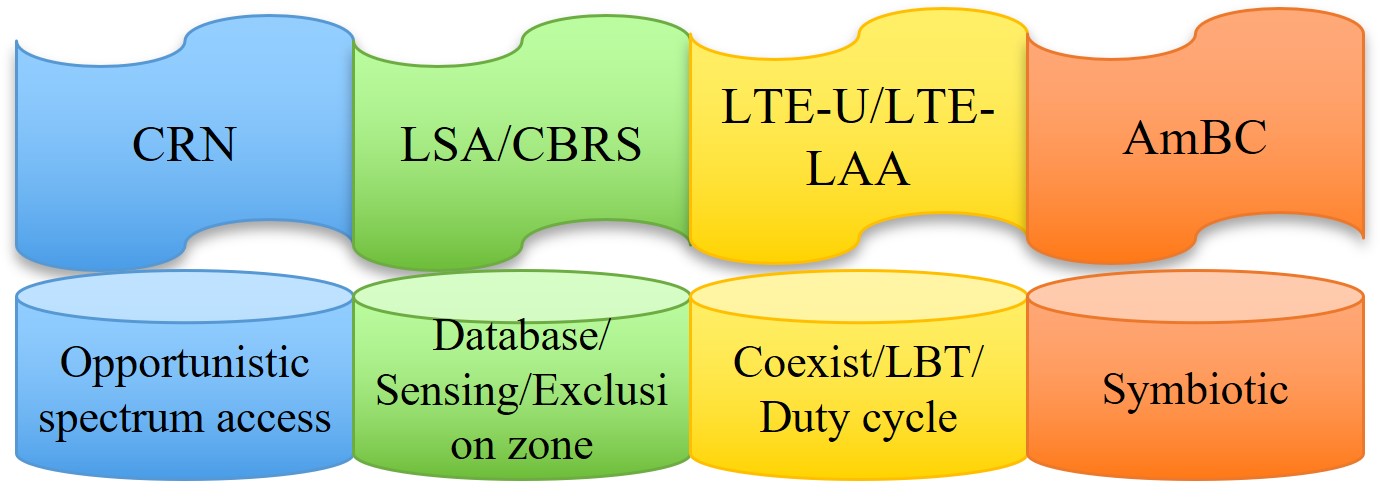}
	\caption{{Spectrum sharing paradigm.\label{ssparadigm}}}
\end{figure}

In summary, as shown in Fig. \ref{ssparadigm}, spectrum sharing originated from the concept of opportunistic access. Database-supported access frameworks on a specific licensed frequency band were then developed to connect new users to the unused licensed band without degrading the performance of IUs to improve the SE. As the number of devices as well as the demand for network capacity has increased, the expansion of licensed band services to the unlicensed band has been proposed. LTE-U provides a view of how to use the unlicensed band to improve the licensed users' performance. Finally, symbiotic schemes such as AmBC help researchers deal with the massive growth of IoT devices that normally have power and resource limitations, providing a new paradigm for spectrum sharing. Although these research studies share some overlapping features such as sensing and access control, they each have their distinctive focuses. 5G has a very broad technical scope and needs to address a variety of device communication problems. Therefore, investigating SS issues under different frameworks can provide very instrumental insights for future communication development.

Nevertheless, to fully exploit the great potentials of SS, users need to interact with the complex and dynamic radio environment in order to gain timely channel access and control the interference. Relying only on traditional radio access technologies may not be sufficient to tackle such complexity. 
With the advancement of computing technologies such as GPU and algorithm development, ML has garnered tremendous interest and recently demonstrated astonishing potential for tackling large-scale, highly dynamic, very complicated problems that traditional techniques cannot easily handle.
ML algorithms have gained advantages in processing, classification, decision-making, recognition, and other problems \cite{ml01}.

SS frameworks naturally share common features with ML, making the combination of ML with SS networks very appealing. For all the frameworks mentioned above,  users/coordinators in the SS network need to observe the spectrum resource usage and make corresponding decisions in accordance with the three main conditions for intelligence, i.e., perception, learning, and reasoning \cite{ml01}. In ML, the intelligent agent first senses the surrounding environment and internal states through perception to obtain information. It further transforms that information into knowledge by using different classification methodologies and generalizes the hypothesis. Based on the obtained knowledge, it then achieves certain goals through reasoning.

The combination of SS with ML techniques has the potential to adaptively tackle complicated and dynamic allocation and classification problems such as channel selection, interference control, and resource allocation.
Applying ML to different SS frameworks has become a research hotspot and a promising frontier in the field of future communications. Towards that end,  this study aims to provide a timely and comprehensive survey in this up-and-coming field.

\subsection{Security in SS Systems}
The development of SS techniques will help relieve spectrum scarcity. However, due to the dynamic access of spectrum resources by a variety of users, SS systems can be exposed to malicious attackers.
In the first place, the lack of ownership of the spectrum leaves unlicensed users highly susceptible to malicious attacks. Therefore, it is hard to protect their opportunistic spectrum access from adversaries. {In the second, the dynamic spectrum availability and distributed network structures make it challenging to implement adequate security countermeasures.} Moreover, in some SS systems, PUs may contain sensitive information, which can be effortlessly obtained by malicious SUs during the SS process. Finally, new technologies such as ML may also be exploited by attackers with even more complicated and unpredictable attacks \cite{jam02}.
Clearly, security issues are of great concern and impose unique challenges in the SS network.

According to \cite{sssi01}, security requirements in most of the SS scenarios include confidentiality, integrity, availability, authentication, non-repudiation, compliance, access control, and privacy. 
Confidentiality means sensitive information should not be disclosed to unauthorized users, especially in database-assisted SS systems. 
Integrity ensures that information communicated among users is protected from malicious alteration, insertion, deletion, or replay.
Availability assures users access to the spectrum/database when it is required.
Authentication requires that the users should be able to establish and verify their identity.
Non-repudiation means users should be able to deny having received/sent a message or to deny having accessed the spectrum at a specified location and time.
Compliance means the network should be able to detect non-compliant behavior that results in harmful interference. 
Access control indicates that users should not access the spectrum/database without credentials.
Privacy means users' sensitive/private information should be protected.

Diverse security threats in different network layers can prevent the SS system from meeting the above requirements. In this work, we will mainly focus on the threats and mitigation strategies in the physical layer of the SS network. We investigate works related to two classical spectrum sensing attacks in the SS network, i.e., PUE attacks and SSDF attacks, which aim to disturb the spectrum observation and users' access to the system. { We also studied methods of preventing two attacks that commonly exist in wireless communication networks, i.e., jamming attacks and eavesdropping attacks.} The special features of the SS network provide new defense solutions for these common attacks. Since PUs need to open their exclusive license spectrum to coexist with multi-type users in some SS frameworks, we also investigated privacy issues and corresponding countermeasures.

To further enhance the security performance of the SS system, ML has become an important part of security and privacy protections in various applications. 
ML is a powerful tool for data exploration and can distinguish normal and abnormal behaviors based on how devices in the SS system interact with each other during spectrum access. The behavioral data of each component in the SS network can be collected and analyzed to determine normal patterns of interaction, thereby allowing the system to identify malicious behaviors early on. Furthermore, ML can also be used to intelligently predict new attacks, which often are the mutations of previous attacks by exploring the existing records. Consequently, SS networks must transition from merely facilitating secure communication to security-based intelligence enabled by ML for effective and secure systems.
The state-of-the-art learning-based security solutions for SS systems will also be comprehensively reviewed in this paper.

\subsection{Motivation and Contribution}
\begin{table*}[h]
   \caption{Related surveys on ML based SS network and security issues. \label{table1} \centering}
    \centering
 
\begin{tabular}{ |p{3cm}||p{1cm}|p{1cm}|p{1cm}|p{1cm}|p{1cm}|p{1cm}|p{1cm}|p{1cm}|p{1cm}|p{1cm}|  }
 \hline
Reference& CRN &LSA/SAS&LTE-U/LAA& AmBC &ML Based&PUE&SSDF&Jamming&Eavesdrop&Privacy\\
 \hline
 R. H. Tehrani et al. \cite{licesy1}   & No    &Yes&   Yes & No    &No&   No & No    &No&   No&No\\
  \hline
M. Massaro et al. \cite{sshs03}& No    &Yes&   Yes & No    &No&   No & No    &No&   No&No\\
 \hline
M. Höyhtyä et al. \cite{sshs033} & No    &Yes &   No & No    &Yes &No & No    &No&   No&No\\
 \hline
N. Van Huynh et al. \cite{amb01}     & Yes    &No&   No & Yes    &No&   No & No    &No&   No&No\\
 \hline
M. Bkassiny et al. \cite{ml01}& Yes    &No&   No & No    &Yes&   No & No    &No&   No&No\\
 \hline
I. F. Akyildiz et al. \cite{cr001}& Yes    &No&   No & No    &No&   No & No    &No&   No&No\\
 \hline
F. Hu et al.  \cite{crnewsy00}& Yes    &No&   No & No    &No&   No & No    &No&   No&No\\
  \hline
L. Zhang et al. \cite{sstr002} & Yes    &No&   No & Yes    &No&   No & No    &No&   No&No\\
 \hline
L. Zhang et al. \cite{sstr003}& Yes    &No&   Yes & No    &No&   No & No    &No&   No&No\\
 \hline
M. A. Hossain et al. \cite{cr00}& Yes    &No&   No & No    &Yes&   Yes & Yes    &No&   No&Yes\\
 \hline
 A. Kaur et al. \cite{crml001}& Yes    &No&   No & No    &Yes&   No & No    &No&   No&No\\
 \hline
J. Park et al. \cite{sssi01}& Yes    &Yes&   No & No    &No&   Yes & Yes    &Yes&   No&Yes\\
 \hline
A. G. Fragkiadakis et al.\cite{ssstr01}& Yes    &No&   No & No    &No&   Yes & Yes    &Yes&   No&No\\
 \hline
] V. Ramani et al. \cite{ssct002} & Yes    &Yes&   No & No    &No&   Yes & Yes    &Yes&   Yes&No\\
 \hline
  Our survey & Yes    &Yes&   Yes & Yes    &Yes&   Yes & Yes    &Yes&   Yes&Yes\\
 \hline
\end{tabular}

\end{table*}

There have been several literature surveys that focus on SS techniques and security issues \cite{cr001, crnewsy00,licesy1, sshs03, sshs033, amb01, sstr002, sstr003, ml01, cr00, crml001, sssi01, ssstr01, ssct002}. However, most of those works only covered one framework of the SS networks.
In \cite{licesy1}, the authors studied the main concepts of dynamic SS and major challenges associated with sharing of licensed bands.
\cite{sshs03} discussed two data-based SS frameworks: LSA and CBRS  and the existing work related to their management approach.
\cite{sshs033} presented existing work in spectrum prediction and learning, summarized the related applications, techniques, main metrics, as well as computational complexity, and provided practical examples.  
\cite{amb01} presented fundamentals of backscatter communications, the general architecture, advantages, and solutions to existing issues and limitations of AmBC systems. 
\cite{cr001} provided an overview of the cognitive radio technology and the architecture in the next-generation network that included different layer and cross-layer protocols.
\cite{crnewsy00} presented a comprehensive survey of CR technology and focused on the research in progress on the full SS with regard to several scenarios. Key enabling technologies were presented in terms of full-duplex spectrum sensing, spectrum-database-based spectrum sensing, auction-based spectrum allocation, and carrier aggregation-based spectrum access.

\cite{cr001, crnewsy00, licesy1, sshs03, sshs033, amb01} considered only one framework of SS and were far from comprehensive. \cite{sstr002,sstr003} investigated the works of different SS frameworks, \cite{sstr002} provided a survey of prevalent IoT technologies employed within the licensed cellular spectrum and unlicensed spectrum. Various SS solutions, including shared spectrum, interference model, and interference management were discussed. 
\cite{sstr003} elaborated advanced techniques for SS such as CR device-to-device communication, in-band full-duplex communication, non-orthogonal multiple access, and LTE-U. For each technique, they presented the basic principles and research methodology of the emerging technology.

However, most of the above surveys only deal with the research progress of traditional SS methods, which were based on relatively simple static settings. Such settings have very little bearing on reality, especially in 5G communication systems. The explosive growth of user equipment, higher data rate requirements, and shorter time delay response requirements, etc., have greatly escalated the complexity of the problem. It is especially challenging when considering these changes in SS networks since systems need to solve the coordination problem of dynamic spectrum access (DSA) and the interference control problem under the coexistence mode. Thus, it is necessary to explore the development of ML in order to design new solutions to those problems in SS systems.

\cite{ml01,cr00,crml001} reviewed ML-assisted SS networks.
In \cite{ml01},  various learning problems in CRN were studied and classified as supervised or unsupervised. Possible solutions were also presented.
An overview of ML, CRN, vehicular ad hoc network (VANET), and CR-VANET was presented in\cite{cr00}, including their architectures, functions, challenges, and open issues. The applications and roles of ML methods in CR-VANET scenarios were reviewed.
In \cite{crml001}, the authors presented a classification and survey of various ML techniques for intelligent spectrum management with their paradigms of optimization for CRNs. 

The above works were limited to typical SS networks such as CRN and did not include new SS techniques. 
Furthermore, research on security in SS networks, especially in systems associated with emerging technologies, were not investigated. 

In \cite{sssi01}, the authors reviewed and classified the critical security and privacy threats that impact SS and described representative examples for each threat category. The threat countermeasures and enforcement techniques were also discussed in the context of two different approaches: ex-ante (preventive) and ex-post (punitive) enforcement.
The authors in \cite{ssstr01} gave an overview of the security threats and challenges that endanger CRN along with the latest technology used to detect corresponding attacks. 
The paper \cite{ssct002} surveyed state-of-the-art research on spectrum sensing and security threats in CRN. Issues related to spectrum handoffs in CRN were also considered.

Nevertheless, these works did not include the protection solutions for security issues in SS networks based on ML algorithms. Application of ML in security countermeasures will produce effective solutions as well as potentially introduce challenges never encountered by researchers before.
Till now, a literature survey that covers advanced SS frameworks with ML techniques in an intelligent manner or considers the security issues in such a context does not exist. 
This article systematically summarizes existing SS technologies based on ML, provides the state of current progress, and paves the way for future development. It provides a timely survey on the most recent works that use ML to solve problems faced in four SS frameworks and corresponding security issues.
{A comparison of existing surveys and our work is presented in table \ref{table1}.}

To summarize, the main contributions of this article are as follows:
\begin{itemize}
{
    \item[(1)]A back review of the development history of SS is given, and  an extensive investigation  of various types of SS frameworks, i.e., CRN, LSA/CBRS, LTE-U/LTE-LAA, and AmBC, is presented. The similarities and differences among these frameworks are analyzed and discussed. Based on their specific application scenarios, the categorizations and descriptions of state-of-the-art ML-based works are comprehensively provided.
    \item[(2)]Specific security issues for spectrum access like PUE and SSDF attacks are discussed. Security threats for wireless communication such as jamming and eavesdropping attacks are also analyzed in the context of SS. Privacy protections for PUs,  SUs, as well as ML algorithms, are also presented. An extensive survey of state-of-the-art ML-based countermeasures in the SS system is included.
    \item[(3)] To facilitate future research, existing challenges and possible research directions for the combination of SS and ML as well as potential security challenges are discussed. 
    }
\end{itemize}

The rest of the paper is organized as follows. In Section II, we present the ML-based technologies for SS. In Section III, SS security issues in the physical layer are elaborated. State-of-the-art countermeasures for security issues are articulated in Section IV. Open issues for future research are discussed in Section V. The paper is concluded in Section VI.

%
\begin{figure}[h]
	\centering
	\includegraphics[width=3.5in]{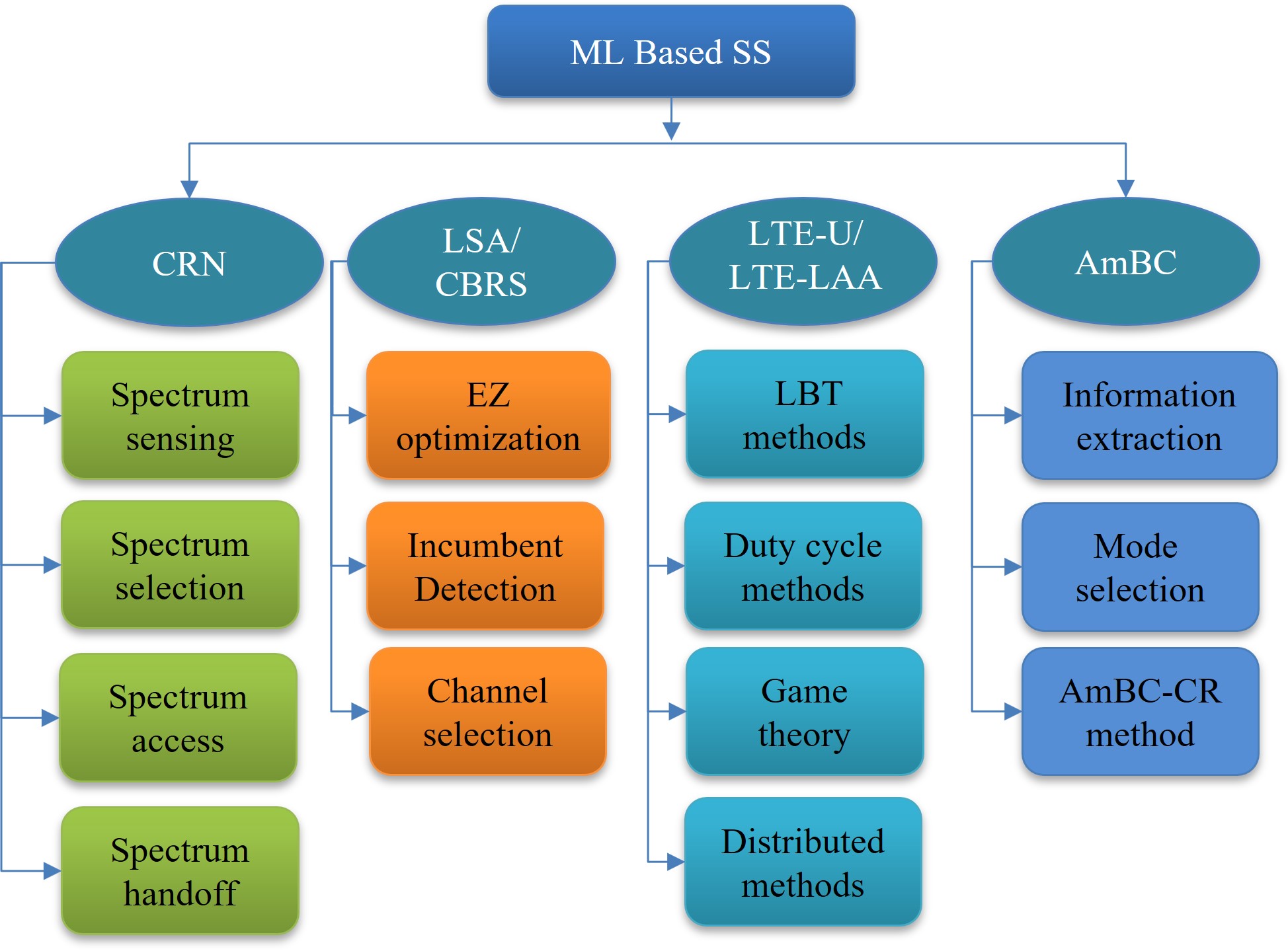}	
	\caption{ {ML-based Methodologies for SS.\label{ml-ss1}}}
\end{figure}


\section{ML-based Methodologies for SS}

In this section, key processes in the CRN network are first introduced, which are shared by most spectrum frameworks. Summarizing the roles that ML plays in these processes can help us to better understand the combination of ML and other frameworks. It also allows us to better assess the security risks in existing SS frameworks with a comprehensive understanding of the mechanism behind each SS technique. The special problems faced by database-assisted SS frameworks (such as LSA and CBRS) are then investigated, i.e., how to protect IUs when unlicensed users are introduced into licensed spectrum bands. Next, a discussion of the application of ML to the coexistence of licensed LTE systems and unlicensed WiFi systems in unlicensed frequency bands is presented. Finally, a comprehensive study of the AmBC system using ML is conducted to gain insight into the symbiosis-based SS framework as well as the benefits of the combination of AmBC and CRN. The content structure is as illustrated in Fig. \ref {ml-ss1}.


\subsection{ML Based CRN}
\begin{figure}[h]
	\centering
	\includegraphics[width=3.3in]{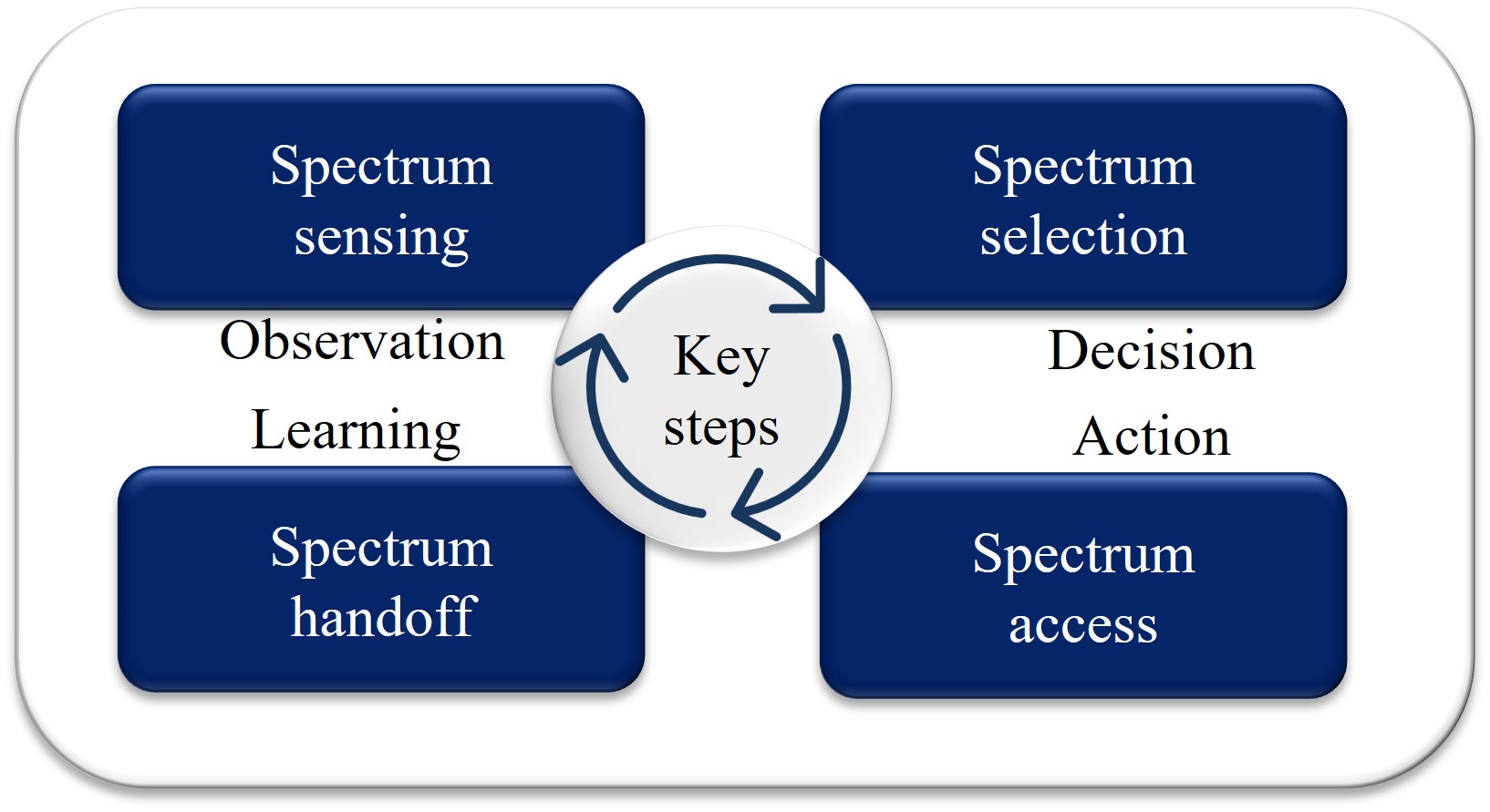}
	\caption{ {Key steps in CRN.\label{crnkeystep}}}
\end{figure}

In CRN, unlicensed SUs need to identify the vacant or unoccupied licensed frequency band (spectrum hole) owned by licensed PUs \cite{cr00}. After spotting the spectrum hole, SUs can access it without visibly interfering with any PU. If a PU's activity reappears, the SUs must vacate the spectrum immediately. 
This dynamic and uncertain environment creates unique and complex challenges within the CRN. However, ML algorithms are very effective in dealing with such challenges and can help improve system performance.

As shown in Fig. \ref{crnkeystep}, the major steps in CRN can be summarized as spectrum sensing, spectrum selection, spectrum access, and spectrum handoff  \cite{cr001}. The CR agent first uses the sensing function to monitor the unused spectrum and search for possible access opportunities for SUs. Based on the sensing results, the spectrum selection function helps SUs select the best available channels and the spectrum access mechanisms provide fair spectrum scheduling among vying SUs. Since a channel must be vacated when the PU reappears, the corresponding SU must perform a spectrum handoff function to switch to another available channel or wait until the channel becomes idle again. It is worth noting that most of the existing vertical SS approaches adopted these four steps in their frameworks.

\subsubsection{Spectrum Sensing}
Before an SU accesses the licensed channel, it needs to first observe and measure the state of the spectral occupancy (i.e., idle/busy) by performing spectrum sensing. During this procedure, the SU needs to distinguish the signal of PUs from background noise and interference.  As such, spectrum sensing can be formed as a classification problem. 

\begin{figure}[h]
	\centering
	\includegraphics[width=3.5in]{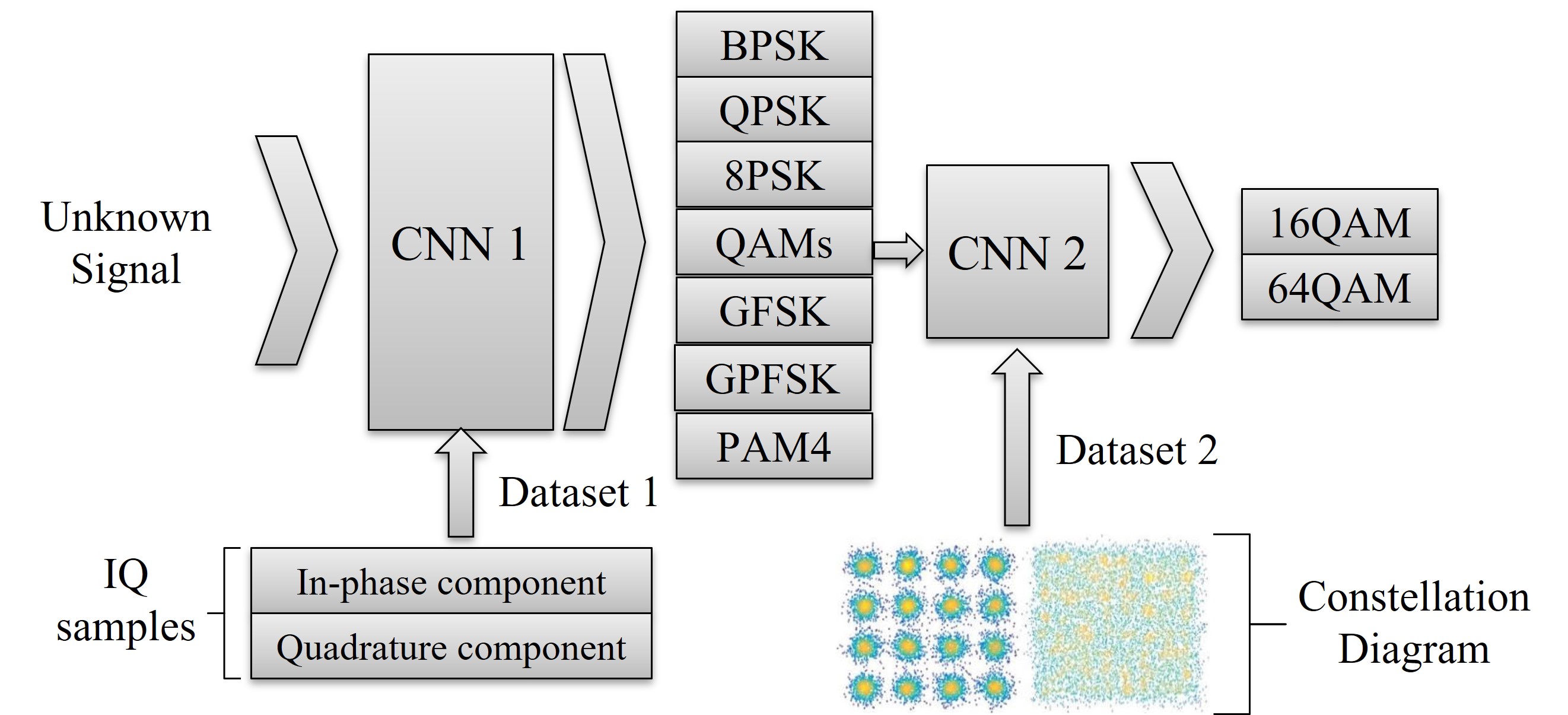}
	\caption{ {Architecture of the DL Automatic modulation recognition system \cite{cr011}.\label{sense1}}}
\end{figure}

Automatic modulation recognition is a keystone of CR adaptive modulation and demodulation capabilities to sense and learn environments and make corresponding adjustments. Automatic modulation recognition can be deemed equivalent to a classification problem, and deep learning (DL) achieves outstanding performance in various classification tasks. Several existing works investigate the combination of DL with automatic modulation recognition in CRNs.
The authors in \cite{cr011} proposed a DL-based framework to achieve accurate automatic modulation recognition. {As shown in Fig. \ref{sense1}, the framework combines two Convolutional Neural Networks (CNNs) that are trained using different datasets.} CNN 1 is trained on samples composed of in-phase and quadrature component signals to distinguish modulation modes that are relatively easy to identify. {
CNN 2 based on constellation diagrams is designed to recognize modulation modes that are difficult to distinguish in CNN 1, such as 16 Quadratic Amplitude Modulation (QAM) and 64 QAM.} This framework demonstrated an outstanding ability to classify QAM signals even in scenarios with a low signal-to-noise ratio.
 In \cite{cr012}, a DL-based automated modulation classification method that employed Spectral Correlation Function (SCF) was proposed. Deep Belief Network (DBN) was applied for pattern recognition and classification. 
By using noise-resilient SCF signatures and DBN that are effective in learning complex patterns, the proposed method can achieve high accuracy in modulation detection and classification even in the presence of environmental noises. {The efficiency of the proposed method was verified in classifying 4FSK, 16QAM, BPSK, QPSK, and Orthogonal Frequency Division Multiple (OFDM) modulation based on various environments settings.}

ML provides effective tools for automating CR functionalities by reliably extracting and learning intrinsic spectrum dynamics. However, there are two critical challenges. First, ML requires a significant amount of training data to capture complex channel and emitter characteristics and train the algorithm of classifiers. Second, the training data that has been identified for one spectrum environment cannot be used for another, especially when channel and emitter conditions change \cite{cr1202}.
To address these challenges, various robust spectrum sensing mechanisms have been developed.
A new approach to training data augmentation and domain adaptation was presented in \cite{cr1202}. {A Generative Adversarial Network (GAN) with DL structures was employed to generate additional synthetic training data to improve classifier accuracy and adapt training data to spectrum dynamics.} This approach can be used to perform spectrum sensing when only limited training data is available and no knowledge of spectrum statistics is assumed. 
Another robust spectrum sensing framework based on DL was proposed in \cite{cr010}. The received signals at the SU's receiver were filtered, sampled, and then directly fed into a CNN. 
{To improve the adaptive ability of the classifier, Transfer Learning (TL) was incorporated into the framework to improve robustness.}

Besides improving the accuracy and robustness of spectrum sensing, another substantial sensing performance improvement comes from using ML to help the SU make efficient decisions regarding which channel to sense and when or how often to sense.

The prediction ability can enable SUs to perform spectrum sensing in a more efficient manner. By enabling SUs to determine the channel selection for data transmission and predicting the period of channel idle status, sensing time can be significantly reduced. {Therefore, the authors in \cite{cr09} proposed an ML-based method that employed a Reinforcement Learning (RL) algorithm for channel selection and a Bayesian algorithm to determine the length of time for which sensing operation can be skipped.} It was shown that the proposed method could effectively reduce the sensing operations while keeping interference with PUs at an acceptable level. This work also showed that by skipping unnecessary sensing, SUs can save more energy and achieve higher throughput by spending the saved sensing time for transmission.
{A Hidden Markov Model (HMM) based Cooperative Spectrum Sensing (CSS) method was proposed in \cite{cr08} to predict the status of the network environment.} First, the concept of an Interference Zone (IZ) was introduced to indicate the presence of PUs. Then, by combing the sensing results from SUs located in different IZs, the Fusion Center employed a fusion rule for modeling specific HMM. Moreover,  the system adopted a Baum-Welch (BW) algorithm to estimate the parameters of the HMM-based past spectrum sensing results. The estimated parameters were then passed to a forward algorithm to predict the activity of PUs.  Finally, SUs were classified into two categories according to the prediction results, i.e., Interfered by PU (IP) and Not Interfered by PU (NIP). SUs marked as IP do not need to perform spectrum sensing to avoid unnecessary energy consumption.


\subsubsection{Spectrum Selection}
After the system receives spectrum sensing results, the spectrum selection is performed to capture the best available spectrum to meet user needs. As a decision-making problem, it requires the system to adaptively capture the optimal choice based on observations of the environment. RL algorithms are appealing tools for designing systems that need to perform adaptive decision-making. In RL, a learner takes actions by trial-error and learns the action patterns suitable for various situations based on the rewards obtained from these actions. Exploration actions are selected in situations even when the knowledge about the environment is uncertain. This mechanism fits into the spectrum selection problems.

\begin{figure}[h]
	\centering
	\includegraphics[width=3.5in]{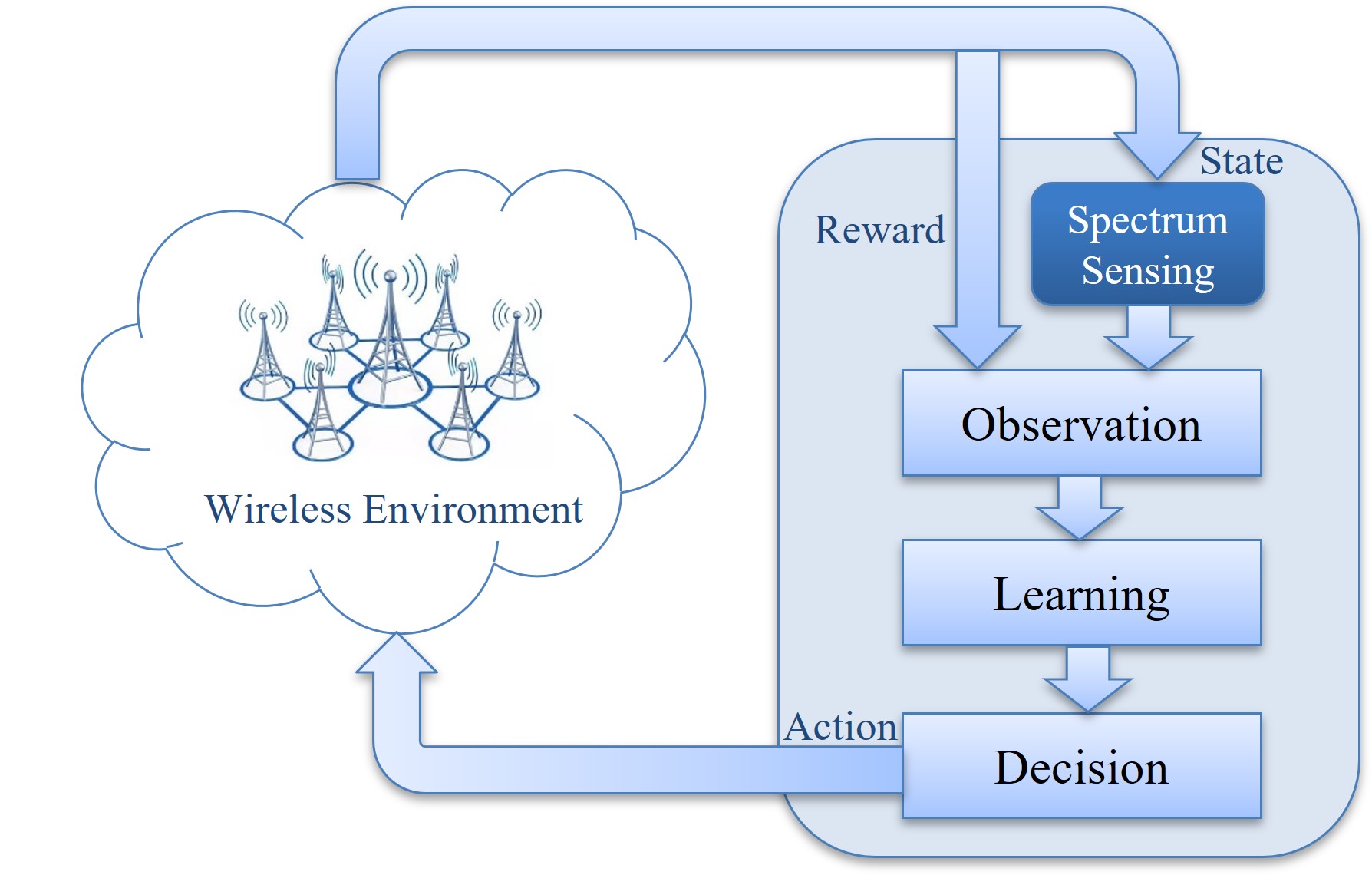}
	\caption{ {The reinforcement learning cycle.\label{rlcycle}}}
\end{figure}

As shown in Fig.\ref{rlcycle}, at the beginning of the RL cycle, the agent receives a full or partial observation of current states and the corresponding reward. Combining those states and rewards, the policy is updated by each agent during the learning stage. Then the agent performs a certain selection action based on the updated policy at the decision stage.
With RL, CRN can be modeled as a distributed self-organized multi-agent system in which each SU or agent performs spectrum selection by efficiently interacting with the environment through a learning policy. In this approach, other SUs' decisions can be considered as a part of the responses of the environment for each SU.

A distributed Learning Automata (LA) based spectrum selection scheme was studied in \cite{cr0801}. It aimed to enable the SUs to sense the RF environment intelligently and to learn its different responses. On the other hand, the activity information of PUs and other SUs was not available, and different SUs could not exchange their information. The self-organized SUs performed the channel selection as the action and received the corresponding response indicating how favorable the action was. Based on the response, SUs could determine the optimal spectrum selection to achieve a lower transmission delay and lower interference to PUs and other SUs.

Two Q-learning based spectrum assignment methods were developed in \cite{cr04}. The independent Q-learning based scheme was designed for a case in which the SUs could not exchange information, while the collaborative Q-learning based scheme was designed for a case in which information could be exchanged among SUs. It was shown that the collaborative Q-learning based assignment method performed better than the independent Q-learning based method.

Most cases fell short of the perfect Channel State Information (CSI), and imperfect CSI can degrade the performance of system spectrum selection.
To overcome this challenge, an RL-based robust decentralized multi-agent resource allocation scheme was proposed in \cite{cr0802}. They introduced cloud computing to enlarge storage space, reduce operating expenditures, and enhance the flexibility of cooperation. A cooperative framework in a multi-agent system was then developed to improve the performance of their proposed scheme in terms of network capacity, outage probability, and convergence speed.


To better manage users in a CRN, the clustering operation organizes SUs into logical groups based on their common features. Clustering can provide network scalability, spectrum stability, and fulfill cooperative tasks.  
In each cluster, one SU can either act as a leader (or cluster head) that manages essential CR operations, such as channel sensing and routing, or as a member node that associates itself with the cluster head.
Cluster size represents the number of nodes in a cluster, and affects various performance metrics. In the CRN, the cluster size adjustment and cluster head selection can significantly impact system performance.

In \cite{cr07}, the authors proposed a first-of-its-kind cluster size adjustment scheme based on RL. The proposed scheme adapts the cluster size according to the number of white spaces to improve network scalability and cluster stability. It was shown that their proposed scheme improved network scalability by creating larger clusters and improved cluster stability by reducing the number of re-clusterings (the number of cluster splits) and clustering overhead while reducing interference between licensed and unlicensed users in CRNs.

Smart malicious SUs can attack the cluster heads to mislead them into inappropriately adjusting the cluster size. To defend against this attack, an RL-based trust model was proposed in \cite{cr05} to improve traditional budget-based cluster size adjustment schemes. The proposed method enables the cluster head to observe and learn about the behaviors of its SU member nodes and to revoke the membership of malicious SUs  in order to alleviate the effects of collaborative intelligent attacks while adjusting the cluster size dynamically according to the availability of white spaces.

{Using Q-value to evaluate the channel quality in Cluster-based CR Ad-Hoc Networks (CRAHN),} a Q-learning-based cluster formation mechanism was studied in \cite{cr06}. Channel quality, residual energy, and network conditions were jointly considered to form a distributed cluster network. All the nodes built their neighboring topology by exchanging the channel status and neighbor list information. Each node then selected the optimal cluster head candidate. Distributed cluster head selections, optimum common active data channel decisions, and gateway node selection procedures were presented. It was shown that the proposed approach could extend the network lifetime and enhance reachability.


\subsubsection{Spectrum Access}
One important question in CRNs spectrum access is how to assign limited resources, such as available spectrum channels and transmit powers, to maximize the system throughput and efficiency. Numerous related works based on RL \cite{cr01, cr02, cr03}, DL\cite{cr013, cr014}, and Deep RL (DRL)\cite{cr016, cr018} have been carried out.


An RL-based resource allocation approach entitled Q-Learning and State-Action-Reward-State-Action (SARSA) was proposed in \cite{cr01} that mitigated interference without the requirements of the network model information. Users in this method act as multiple agents and cooperate in a decentralized manner. A stochastic dynamic algorithm was formed to determine the best resource allocation strategy. It was shown that the energy efficiency could be significantly improved by the proposed approach without sacrificing user QoS.

An energy harvesting enabled CRN was investigated in \cite{cr02}. To achieve higher throughput, a harvest-or-transmit policy for SUs transmit power optimization was proposed. A Q-learning based online policy was developed first to deal with the underlying Markov process without any prior knowledge. An infinite horizon stochastic dynamic programming-based optimal online policy was then proposed by assuming that the full statistical knowledge of the governing Markov process was known. Finally, a generalized Benders decomposition algorithm based offline policy was given, where the energy arrivals and channel states information were known before all transmitters for a given time deadline.

{By combining Multi-Armed Bandit (MAB) and matching theory, the ML-assisted Opportunistic Spectrum Access (OSA) approach was developed in \cite{cr03}.} A single SU case was first considered without the volatility of channel availability information. Next, the upper confidence bound algorithm-based Occurrence-Aware OSA (OA-OSA) framework was designed to achieve the long-term optimal network throughput performance and the trade-off between exploration and exploitation. The OA-OSA was then extended to the multi-SU scenario with channel access competitions by integrating the Gale-Shapley algorithm.


In \cite{cr013}, by considering the OFDMA-based resource allocation for the underlying SUs, researchers aimed to minimize the weighted sum of the secondary interference power under the constraints of QoS, power consumption, and data rate. {A Damped Three-Dimensional (D3D) Message-Passing Algorithm (MPA) based on DL was proposed, and an analogous back-propagation algorithm was developed to learn the optimal parameters.} A sub-optimal resource allocation method was developed based on a damped two-dimensional MPA to improve computational efficiency.
By considering the EE and SE, as well as Computing Efficiency (CE)  for both PUs and SUs, a DL-based resource allocation algorithm in CRNs was proposed in \cite{cr014} to minimize the weighted sum of the secondary interference power. 
It was shown that the proposed scheme significantly improved both the SE and EE for PUs and SUs.

Insufficient specificity and function approximation can impose some limitations on RL algorithms, but neural networks can be used to compensate for them.
DRL algorithms are capable of combining the process of RL with deep neural networks to approximate the Q action-value function. Compared with conventional RL, DRL can significantly improve learning performance and learning speed.
DRL has attracted a lot of attention in research for solving the problems in CR networks such as resource allocation, spectrum management, and power control.

In \cite{cr016}, the authors presented a DRL-based resource allocation method for CRN to maximize the secondary network performance while meeting the primary link interference constraint. {By adopting a Mean Opinion Score (MOS) as the performance metric, the proposed model seamlessly integrates resource allocations among heterogeneous traffic.} The resource allocation problem was solved by utilizing a Deep Q Network (DQN) algorithm where a neural network approximated the Q action-value function. TL was incorporated into the learning procedure to further improve the learning performance. It was shown that TL reduced the number of iterations for convergence by approximately $25\%$ and $72\%$ compared to the DQN algorithm without utilizing TL or standard Q-learning, respectively.

In \cite{cr018}, the authors considered the problem of SS in a CR system where PU is assumed to update its transmitted power based on a pre-defined power control policy. Neither PU's transmit power information nor its power control strategy were available on the SU side. The objective was to develop a learning-based power control method for the SU to share the common spectrum with the PU. To assist the SU, a set of sensor nodes were spatially deployed to collect the received signal strength information at different locations in the wireless environment. They developed a DRL-based method that allowed the SU to adjust its transmit power intelligently. After a few rounds of interactions with the PU, both users could transmit their own data successfully with the required QoS.


\subsubsection{Spectrum Handoff}
Spectrum handoff is intended to maintain seamless communication during the transition to a better spectrum.
However, enabling spectrum handoff for multimedia applications in a CRN is challenging due to multiple interruptions from PUs, contentions among SUs, and heterogeneous Quality-of-Experience (QoE) requirements. 
Although an SU may not know exactly when the PU  comes back, it always wants to achieve reliable spectrum usage to support the QoS requirements. If the quality of the current channel degrades, the SU can make one of the following three decisions:
\begin{itemize}

\item[(1)] Stay in the same channel and wait for it to become idle again (called stay-and-wait). 
\item[(2)] Stay in the same channel and  adapt to the varying channel conditions (called stay-and-adjust).
\item[(3)] Switch to another channel that meets the QoS requirement (called spectrum handoff).
   
\end{itemize}

In \cite{cr01701}, a learning-based and QoE-driven spectrum handoff scheme was proposed to maximize the multimedia users' satisfaction. A mixed preemptive and non-preemptive resume priority (PRP/NPRP) M/G/1 queueing model was designed for the spectrum usage behaviors of prioritized multimedia applications. The RL-assisted QoE-driven spectrum handoff scheme was developed to maximize the quality of video transmissions in the long term. Their proposed learning scheme could adaptively perform spectrum handoff based on the variation of channel conditions and traffic loads.

To address limitations of PRP/NPRP queuing models, the authors in \cite{cr01702} employed a hybrid queuing model with discretion rules to characterize the SUs' spectrum access priorities. The channel waiting time during spectrum handoff was then calculated according to this hybrid queuing model. The multi-teacher knowledge transfer method was further proposed to accelerate the algorithm, wherein the multiple SUs that already had mature spectrum adaptation strategies could share their knowledge with an inexperienced SU.

A Transfer Actor-Critic Learning (TACT) algorithm was proposed for spectrum handoff management in \cite{cr017}.  A comprehensive reward function that incorporated the Channel Utilization Factor (CUF), Packet Error Rate (PER), Packet Dropping Rate (PDR), and flow throughput was adopted in this method. CUF was evaluated by the spectrum sensing accuracy and channel holding time. The PDR was determined based on the non-preemptive M/G/1 queuing model, and the flow throughput was estimated according to a link-adaptive transmission that utilized the rateless codes. Compared to the myopic and Q-learning-based spectrum selection schemes, the proposed scheme achieved a higher mean opinion score.

\begin{figure}[h]
	\centering
	\includegraphics[width=3.5in]{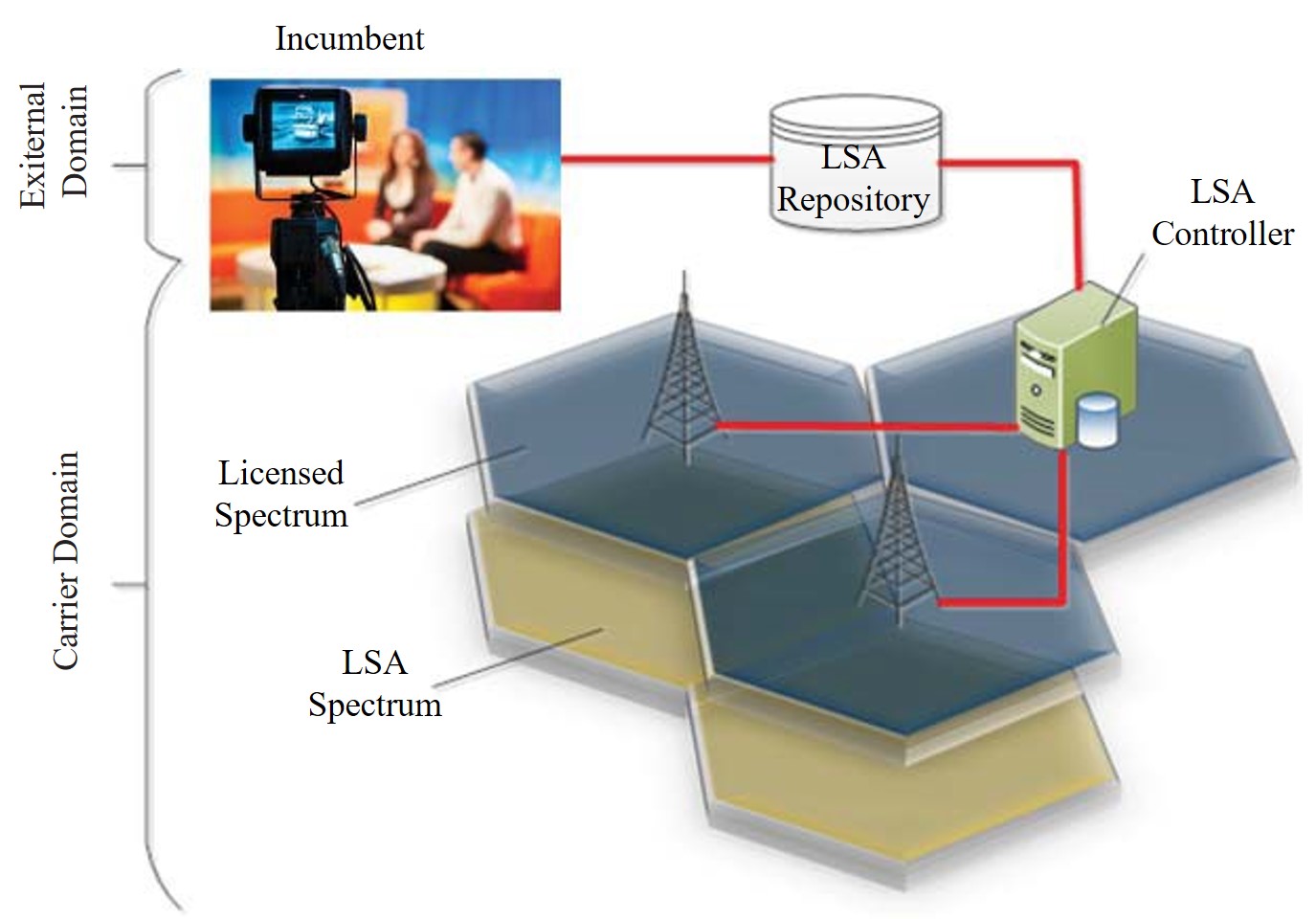}	
	\caption{ {LSA architecture reference model \cite{dbss00}.\label{lsa1}}}
\end{figure}
 \subsection{Database Assisted SS}
The sensing driven OSA system aims to explore the spectrum holes in the unlicensed band, which can be inefficient and unreliable, especially when the number of wireless communication devices increase rapidly. To better serve the secondary system, a specific spectrum band like 3.5 GHz is opened to the public. Database assisted Dynamic SS (DSS) systems such as LSA and SAS have been proposed to better coordinate the users in different systems with various spectrum access priorities.

As shown in Fig. \ref{lsa1}, LSA has two types of users: incumbents and SUs, where incumbents send their spectrum usage information to a database center called LSA Repository \cite{dbss00}. The system then decides whether the SUs can access the spectrum resource with this information or not and no sensing ability is required for those users.

\begin{figure}[h]
	\centering
	\includegraphics[width=3.5in]{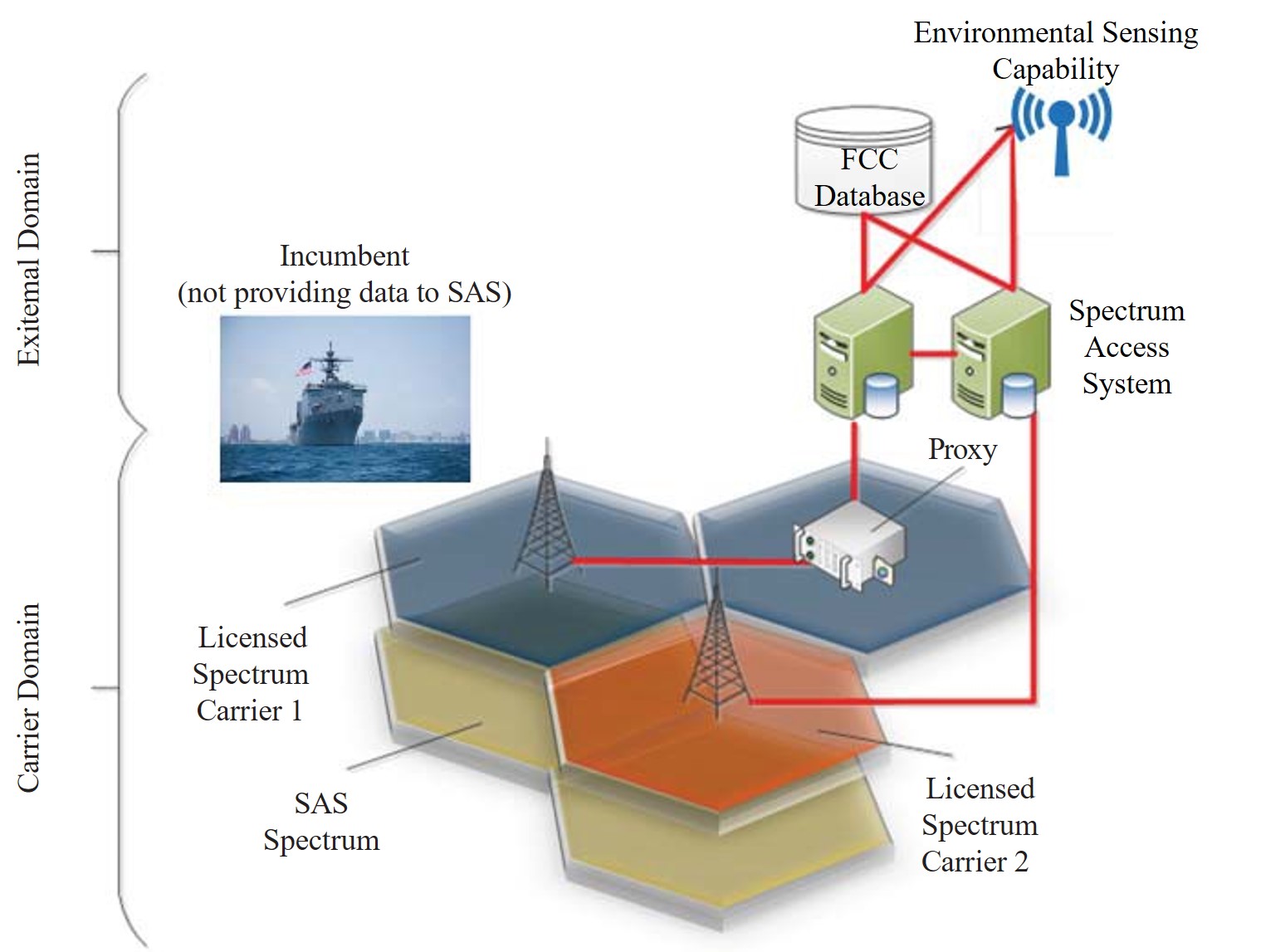}	
	\caption{ {The FCC’s CBRS architecture \cite{dbss00}.\label{sas1}}}
\end{figure}

The SAS system also maintains a similar database for three different types of users as presented in Fig.\ref{sas1}. The difference is that the IUs in the SAS system may have very sensitive information and do not want to offer it to the database \cite{dbss00}. Instead, to protect the IUs, the Exclusion/Protection Zone (EZ) is applied where the SUs (PAL and GAA users) are banned from accessing the spectrum in these areas to prevent them from harmful interference to IUs. Environment sensing capability (ESC)-based incumbent detection also requires users in tier 2 and 3 to perform sensing. The system then decides the spectrum access based on the EZ and ESC nodes' sensing results \cite{gbss01}.

The existing ML-based works for database-assisted SS networks are mainly focused on the EZ adjustment \cite{gbss05, gbss07}, ESC performance improvement \cite{gbss031, gbss03}, and spectrum access coordination \cite{gbss08, gbss09, gbss010, gbss0100}.

\subsubsection{ML based EZ Optimization}
{In a database-driven SS network, even with the SS policy provided by the database,} harmful interference still can occur between a PU and an SU due to unexpected propagation paths. In a common solution, a primary EZ is provided for the PUs that prevents SUs from using the same spectrum in EZ to keep interference at an acceptable level. However, the size of EZ needs to be optimized to efficiently cover the regions where interference may occur and a region's shape is usually not circular.

In \cite{gbss05}, an ML-based framework was developed to deal with the interference by dynamically adjusting the EZ. By considering the propagation characteristics and shadow fading, the framework employed under-sampling and over-sampling schemes to solve an imbalanced data problem which can degrade the estimation accuracy of the appropriate shape of EZ. 
It was shown that their proposed method could significantly reduce the area of the EZ by $54\%$ compared with the fixed circular EZ setting. Furthermore, the proposed sampling scheme could achieve a $1\%$ interference probability with $21\%$ fewer iterations and a $6\%$ smaller area compared with the existing sampling benchmarks.

Using VHF-band radio sensors and the ML technique, an outdoor location estimation scheme of a high-priority DSS system was proposed in \cite{gbss07}. The delay profiles measured in the VHF band were employed to estimate location. The precision of the EZ could then be improved based on the estimated location of the PUs. By using the ARIB STD-T103 system operating in the VHF-band, they measured delay profiles in a mountainous environment in Japan with the Deep Neural Network (DNN). With the trained DNN, the location cluster of the high-priority terminal could be predicted without GPS by simply measuring the delay profile of the PUs. It was shown that their method could significantly improve the total correct localization rate by up to $80.0 \%$.
 

 
 \subsubsection{Incumbent Detection}

According to the FCC, the IUs in the CBRS band include authorized federal users such as U.S. Navy shipborne SPN-43 air traffic control radar operating in the 3550-3700 MHz band, Fixed Satellite Service (space-to-Earth) earth stations operating in the 3600-3650 MHz band, and for a finite period, grandfathered wireless broadband licensees operating in the 3650-3700 MHz band. Due to security restrictions, the SAS cannot access the information of those IUs. To alleviate harmful interference from PAL and GAA users, the Environmental Sensing Capability (ESC) enabled by sensor networks could detect transmissions from the Department of Defense radar systems and transmitted that information to the SAS. The SAS could then assign the spectrum resources to users with different priorities dynamically. However, the single sensor detection lacked precision due to its geolocation, while distributed multiple sensor networks lead to a high information exchange overhead. Moreover, the extreme operational characteristics of incumbent military wireless applications could overwhelm the existing spectrum sensing methods. 
Several studies have sought to address these issues.

An ML-based spectrum sensing system called Federated Incumbent Detection in CBRS (FaIR) was proposed in \cite{gbss031}. Federated learning (FL) was adopted for ESCs to collaborate and train a data-driven ML model for IU detection with minimal communication overhead. Unlike a naive distributed sensing and centralized model framework, their proposed method could exploit the spatial diversity of the ESCs and improve detection performance.

Because the existing ESC-based methods had the potential to incur a high communication overhead and lead to leakage of sensitive information, a compressed sensing (CS)-based FL framework was proposed in \cite{gbss03} for IU detection. To protect privacy, local learning models transmitted updating parameters instead of raw spectrum data to the central server. A Multiple Measurement Vector (MMV) CS model was further adopted to aggregate these parameters. Based on the aggregated parameters, the central server could gain a global learning model and send the global parameters back to local learning models. Their proposed framework could significantly improve communication and training efficiency while guaranteeing detection performance compared with the raw training sample method.


 
 
 \subsubsection{Channel Selection and Transaction}
In the database-assisted SS system, each user needs to choose a proper vacant channel in order to avoid severe interference with others. When economic approaches are adopted to model the SS system, idle channels can be traded as commodities. For this reason, channel selection and spectrum trade problems are critical to the database-assisted SS system, and many works based on game theory have been proposed to solve them.

 In \cite{gbss08}, a database-assisted distributed white-space Access Point (AP) network design was studied. The cooperative channel selection problem was first considered to maximize system throughput, where all APs were owned by one network operator. A distributed channel selection problem was then formed between APs that belonged to different operators, and a non-cooperative state-based game was formulated by considering the mobility of SUs. 
 It was also shown that this algorithm was robust to perturbation from SUs' leaving and entering the system.

 In \cite{gbss09}, a method of idle channels sharing in overlapped licensed areas among PUs and SUs was proposed. Based on supply and demand fluctuations in different areas, SUs were grouped based on their suppliers, and channel transaction quotas were set by PUs for these SU groups accordingly. By applying evolutionary games, the PUs could obtain quotas of Evolutionary Stable Strategy (ESS) to maximize their incomes. Furthermore, a learning process was designed for the PUs to attain the optimal realizable integer quotas.

In \cite{gbss010}, the authors jointly considered geographic location, price-based channel blocks available at PUs, and the demand from SUs to determine channel selling preference, channel selection decisions, and channel prices for PUs and SUs. To maximize the revenue of PUs, a unique quota transaction process was proposed. By applying an evolutionary procedure defined as replicator dynamics, the existence and uniqueness of the evolutionarily stable strategy quota vector of each PU was proved, and the optimal payoff of each PU for selling channels without reservation was determined. In the scenario of selling channels with reservation, channel price prediction was employed to assist decisions in optimal supplies strategies.


\subsection{ML Based LTE-U/LTE-LAA}

LTE-U has emerged as an effective technique for alleviating spectrum scarcity. Using LTE-U along with advanced techniques such as carrier aggregation can boost the performance of existing cellular networks. 
 However, LTE was initially designed to operate in the licensed spectrum exclusively and was not for harmonious coexistence with other possible co-located technologies \cite{lteu010}.
For this reason, introducing LTE into the unlicensed spectrum can cause significant coexistence issues with other well-established unlicensed technologies such as Wi-Fi, IEEE 802.15.4, or Bluetooth. To enable fair spectrum sharing with other technologies operating in the unlicensed spectrum, in particular with Wi-Fi, new coexistence technologies are needed. On the other hand, not much research attention has been given to studying cooperation between the technologies. Networks that participate in a cooperation scheme can exchange information directly or indirectly (via a third-party entity) to improve the efficiency of spectrum usage in a fairway.
 
\begin{figure}[h]
	\centering
	\includegraphics[width=3.6in]{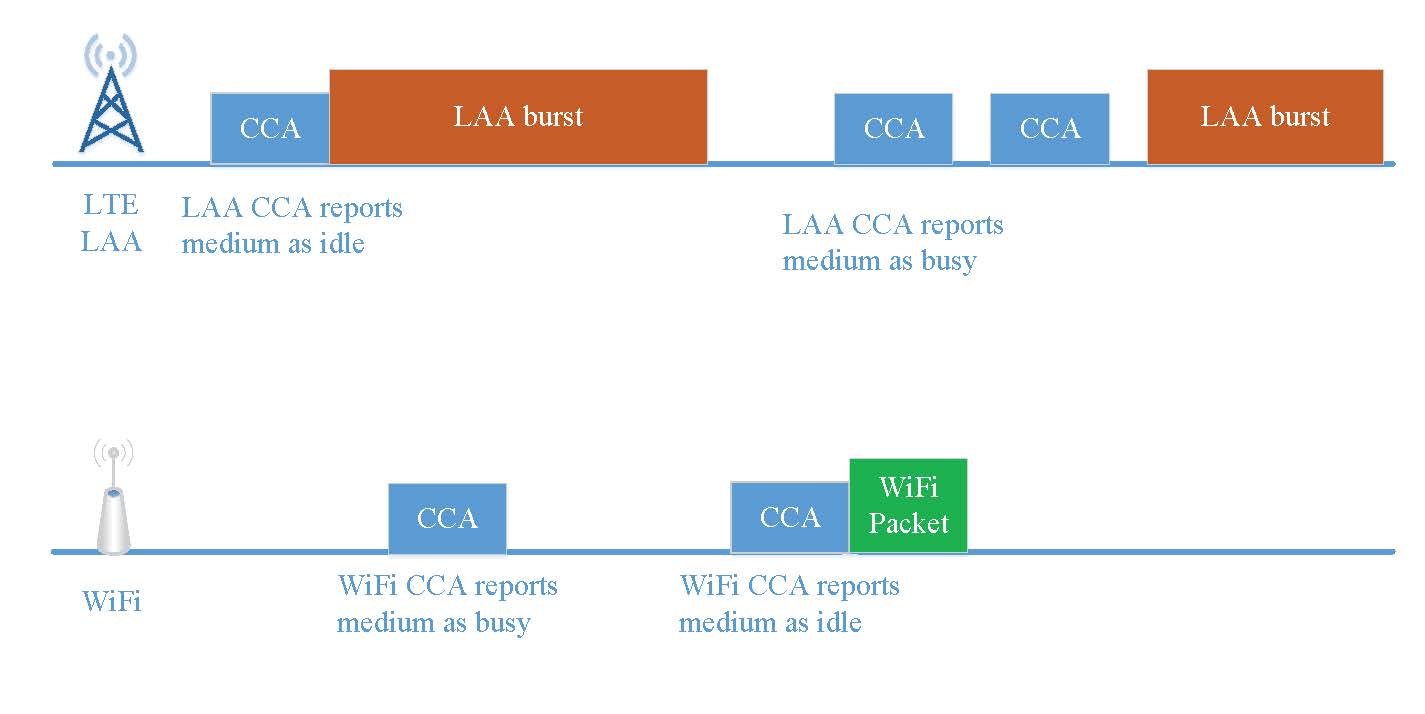}	
	\caption{ {LBT based method.\label{LBT}}}
\end{figure}

To standardize LAA technology in the $5$ GHz spectrum, the Third-Generation Partnership Project (3GPP) standardization group aims to develop a single global framework of LTE in the unlicensed bands. The framework should guarantee that the operation of LTE does not critically affect the performance of WiFi networks. Initially they only considered the downlink operation LTE-A (LTE Advanced) Carrier Aggregation (CA) in the unlicensed band. This was later expanded to include the simultaneously operate downlink and uplink \cite{lteu04}.
The LTE LAA employed a Listen Before Talk (LBT) mechanism to avoid collision and interference between users. 

 LTE-U is another option for operating LTE in an unlicensed spectrum, where LTE base stations exploit transmission gaps to facilitate coexistence with WiFi networks. The development of LTE-U technology is led by the LTE-U Forum, an industry alliance. LTE-U has been designed to operate as an unlicensed LTE in countries where the LBT technique is not mandatory, such as the United States and China. LTE-U defines the operation of primary cells in a licensed band with one or two secondary cells (SCells), every $20$ MHz in the $5$ GHz unlicensed band: U-NII-1 and/or U-NII-3 bands, spanning $5150$–$5250$ MHz and $5725$–$5825$ MHz, respectively \cite{lteu04}. 


In addition to LBT technique \cite{lteu08, lteu011, lteu071} or duty cycle-based methods \cite{lteu04, lteu05, lteu06, lteu010}, the game theory-based SS for LTE with WiFi has also been investigated in \cite{lteu03, lteu02, lteu07, lteu01, lteu012, lteu013, lteu014}, where each system formed its own spectrum allocation decision and negotiated with the others to achieve the best coexistence results.


\subsubsection{ML Based LBT Methods}

According to LTE LAA standards in 3GPP Release 13, the LTE system must perform the LBT procedure (also known as Clear Channel Assessment, CCA) and sense the channel prior to a transmission in the unlicensed spectrum. As shown in Fig. \ref{LBT}, when the channel is sensed to be busy, the LTE system must defer its transmission by performing an exponential backoff. If the channel is sensed to be idle, it performs a transmission burst with a duration from $2-10$ ms, depending on the channel access priority class \cite{lteu0101}.

To exploit the benefits of communications in an unlicensed spectrum using LTE-LAA, a DL approach for the resource allocation of LTE-LAA small base stations (SBSs) was proposed in \cite{lteu08}. The proposed method employs a proactive coexistence mechanism that enables future delay-tolerant LTE-LAA data requests to be served within a given prediction window before their actual arrival time. Therefore, it can improve the utilization of the unlicensed spectrum during off-peak hours while maximizing the total served LTE-LAA workloads. 
To achieve long-term equal-weighted fairness between wireless local area networks and LTE-LAA operators, a non-cooperative game model was formulated where SBSs were modeled as homo egualis agents to predict a future action sequence. The proposed method enables multiple SBSs to proactively perform dynamic channel selection, carrier aggregation, and fractional spectrum access while guaranteeing equal opportunities for existing WiFi networks and other LTE-LAA operators.

In \cite{lteu011}, the computing offloading problem in an LTE-LAA enabled network was investigated, where IoT devices could either process their tasks locally or offload them to an LTE-U base station. 
To maximize the long-term discount reward, an offloading policy optimization problem was formulated by considering both task completion profit and task completion delay. Category 4 channel access scheme (LBT with random back-off and a contention window of variable size) was adopted in the LTE-U network to sense and occupy the unlicensed spectrum resources. 
The DQN algorithm was applied to solve the offloading problem by considering each device's stochastic task arrival process and the Wi-Fi's contention-based random access.

An analytical model was proposed in\cite{lteu071} to evaluate the throughput performance of the Category 4 algorithm agreed in the 3GPP release 13. A semi-branch and bound algorithm was adopted to maximize the aggregate throughput of the LTE-LAA and Wi-Fi networks under the system fairness constraints. RL methods were further introduced to adjust the contention window size for both LTE-LAA and Wi-Fi nodes. Based on the assumption that information is exchangeable between different systems, a cooperative learning algorithm was first developed then extended to a more practical non-cooperative version.

Inspired by the LBT techniques in the LTE-LAA method, the authors in \cite{gbss0100} proposed and compared two LBT based schemes for using the CBRS spectrum. The proposed methods allow the GAA users to opportunistically access the PAL spectrum with acceptable interference. A $50\%$ User Perceived Throughput (UPT) gain was achieved for SUs with a minor decrease in PUs' UPT. Q-learning was further used to adjust the SUs’ opportunistic access by learning the Energy Detection Threshold (EDT) for carrier sensing to combat hidden and exposed nodes problems. As a result, it was shown that SUs could improve their throughput by up to $350\%$ with merely a $4\%$ reduction in PUs’ UPT when adapting the EDT of opportunistically transmitting nodes.


\subsubsection{{ML based Duty Cycle Methods}}

\begin{figure}[h]
	\centering
	\includegraphics[width=3.6in]{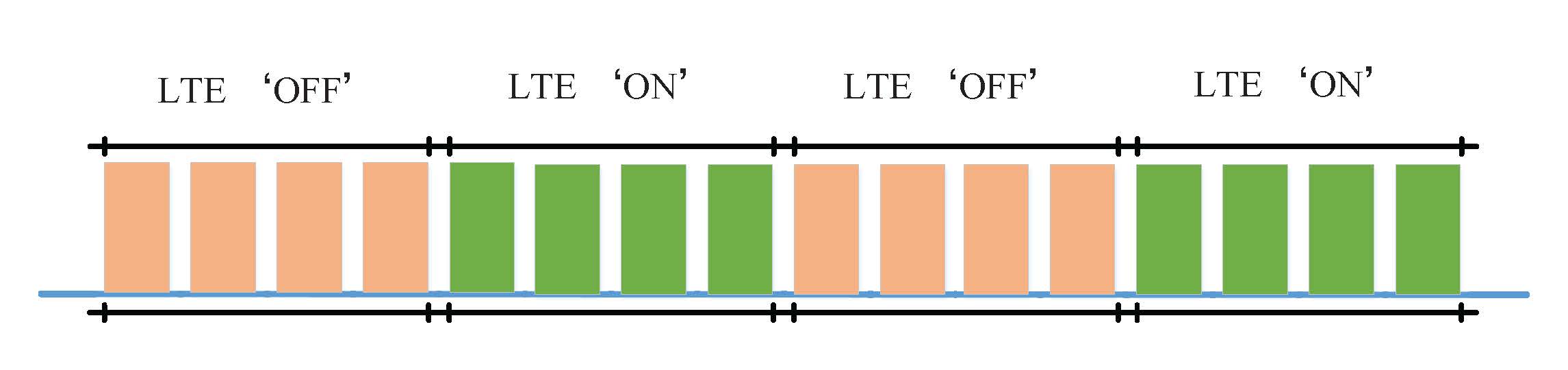}	
	\caption{ {Duty cycle based method.\label{duty}}}
\end{figure}

Carrier Sensing Adaptive Transmission (CSAT) is a technique that can enable coexistence between LTE and Wi-Fi based on minor modifications of the 3GPP LTE Release 10/11/12 Carrier Aggregation protocols. {As shown in Fig. \ref{duty}, CSAT introduces the use of duty cycle periods and divides the time into LTE “ON” and LTE “OFF” slots.} During the LTE “OFF” period, also known as the “mute” period, LTE remains silent, giving other coexisting networks, such as Wi-Fi, the opportunity to transmit. During the LTE “ON” period, LTE accesses the channel without sensing it before transmission. Moreover, CSAT allows short transmission gaps during the LTE “ON” period to allow for latency-sensitive applications, such as VoIP in co-located networks. {In CSAT, eNB senses the medium during a time period ranging from 10 to 100 ms and according to the observed channel utilization (based on the estimated number of Wi-Fi APs) defines the duration of the LTE “ON” and LTE “OFF” periods \cite{lteu010}.}

The existing work of LTE-U mainly focuses on using different RL algorithms to adjust the duty cycle and other network resources to maintain fairness between LTE and WiFi users, as well as to seek for a higher system capacity performance.


To investigate the application of LTE-U technology in the $3.5$ GHz CBRS band, an MAB-based SS technique was developed in \cite{lteu04} for a seamless coexistence with WiFi. Assuming LTE-U to operate as a GAA user, they used MAB to adaptively optimize the duty cycle of LTE-U transmissions. Downlink power control was incorporated to achieve high EE and interference suppression. 
The study showed significant improvement in the aggregate capacity and cell-edge throughput of coexisting LTE-U and WiFi networks for different base station and user densities.


In \cite{lteu05}, a Q-learning based LTE-U and Wi-Fi coexistence algorithm was proposed in multi-channel scenarios. To enable the alternate data transmission in LTE-U and Wi-Fi coexist systems, the algorithm optimized the duty cycle by jointly considering both fairness and overall system performance. The efficacy of the proposed algorithm in improving throughput has been verified with the premise of ensuring fairness.  
A Q-Learning-based approach was proposed in \cite{lteu06}, which enabled the LTE-U BS to dynamically identify and exploit white spaces in the WiFi channels without requiring detailed knowledge of the WiFi system. The proposed approach aims to minimize the latency imposed on WiFi activity by employing carrier sensing at the BS while maximizing use of idle spectral resources for unlicensed LTE. By adaptively adjusting the LTE-U duty cycle to WiFi activity, the proposed algorithm enabled maximal utilization of idle resources for LTE-U transmissions while decreasing the latency imposed on WiFi traffic.

In \cite{lteu010}, a duty-cycle SS framework was first designed that allowed an LTE system to share the spectrum with a WiFi system in the time domain. A DRL-based algorithm was then developed to learn WiFi traffic demands by analyzing WiFi channel status. To maximize the LTE system's throughput and provide sufficient protection to the WiFi system, the LTE system adaptively optimized its transmission time based on learned knowledge. It was shown that the performance of the proposed method could approach the performance of the genie-aided exhaustive search algorithm, which is based on a perfect knowledge of WiFi traffic demands through massive signaling exchanges and is of high computational complexity,in terms of LTE throughput and WiFi protection.


\subsubsection{{Game Theory Based Methods}}
The coexistence of LTE systems and WiFi system can usually be formed as a game theory problem, where each part must compete for the same unlicensed spectrum resource and finally reach an agreement with each other. Some game-theory-based LTE-U works are discussed in this section.

\begin{figure}[h]
	\centering
	\includegraphics[width=3.6in]{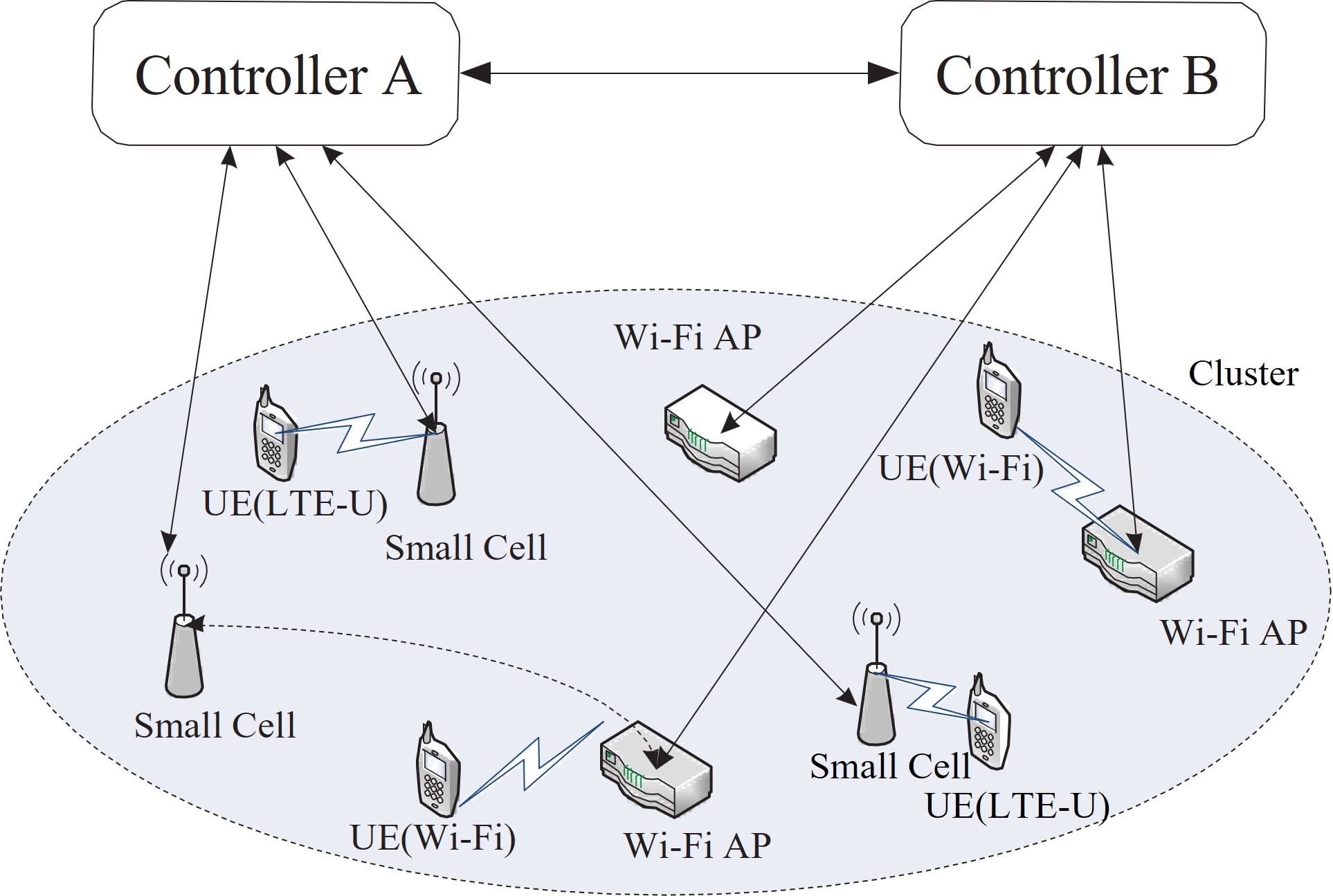}	
	\caption{ {The coexistence of LTE-U and WiFi in an unlicensed spectrum \cite{lteu03}.\label{lteugm}}}
\end{figure}

An SS scheme adapted for the operation of LTE-U and WiFi systems was proposed in \cite{lteu03}. The decision tree learning and repeated game were adopted to optimize unlicensed spectrum resource. As shown in Fig.\ref{lteugm}, the control plane was decoupled from the data plane and provided the system with a great capacity for processing data. Controllers learned the latest dataset in the pool to build decision trees and deduce the network status of the opponent. Repeated games for sharing spectrum resources were then employed to maximize resource utilization in the coexistence system. {An incentive mechanism was further included to increase operators' motivation to share their spectrum resources.}

To maximize the users' satisfaction across the network, an optimization problem was formulated in \cite{lteu02} by jointly investigating spectrum access, power allocation, and user scheduling. A game-theoretic approach and interference graph were employed to solve this problem. The proposed game was proved to have at least one Nash Equilibrium (NE), corresponding to either the globally or locally optimal solution to the original problem. To solve the problem, a concurrent best-response iterative algorithm capable of converging an NE was developed. However, this algorithm could not guarantee the global optimality and a Spatial Adaptive Play Iterative (SAPI) learning algorithm was further developed for the global optimum search.

The authors in \cite{lteu07} studied unlicensed band selection and resource allocation for the LTE-U network with respect to users' QoE, which was measured in MOS. An optimization problem was formulated to maximize LTE users' QoE while protecting the Wi-Fi system. To resolve the multi-player interaction in this spectrum space fairly, a Virtual Coalition Formation Game (VCFG) was used to solve the unlicensed band selection problem. The results of the VCFG defined the optimization problem within each coalition. This problem was then decomposed into a time-sharing problem between the LTE-U and Wi-Fi systems and a resource allocation problem for the LTE-U system. The first sub-problem was solved by the cooperative Kalai-Smorodinsky bargaining solution, whereas the second was solved by the Q-learning algorithm. Finally, VCFG and Q-learning-based resource allocation algorithms were proposed.

To enable the cellular network to use LTE-U with CA to meet the QoS needs of its users while protecting Wi-Fi Access Points (WAPs) for a network with multiple operators, the problem of LTE-U sum-rate maximization was addressed in \cite{lteu01} under user QoS and WAP-LTE-U coexistence constraints. A cooperative Nash Bargaining Game (NBG) and a one-sided matching game were introduced to solve this problem. The NBG deals with the coexistence issue between LTE-U and Wi-Fi systems, and the matching game handles the resources allocation problem in the LTE-U system. {These two games repeat till convergence. This proposed approach outperformed other comparing benchmarks in terms of achieved rate, users' satisfaction, and fairness.} 

\subsubsection{Distributed Algorithm based Methods}

In \cite{lteu012}, the authors investigated the uplink-downlink decoupling resource allocation problem for LTE-U enabled Small Cell Networks (SCNs). By incorporating user association, spectrum allocation, and load balancing, the problem was formulated as a non-cooperative game. An Echo State Network (ESN) based distributed algorithm was developed to address this problem. It was shown that even with limited information on the network's and users' states, the proposed algorithm could help the SBS to choose their optimal resource allocation strategies autonomously. Furthermore, the proposed method could significantly improve the sum rate of the $50$th percentile of users and achieve a $167\%$ increase compared to a Q-learning algorithm. 

In \cite{lteu013} \cite{lteu014}, a cache-enabled Unmanned Aerial Vehicles (UAVs) communication network serving wireless ground users over the LTE-U bands was considered. The problem under investigation was joint caching and resource allocation. By jointly incorporating user association, spectrum allocation, and content caching, a resource allocation problem was formed and a distributed algorithm based on the Liquid State Machine (LSM) was proposed. The proposed LSM algorithm would enable the cloud to predict users' content request distribution with limited information about network and users. The proposed algorithm will also help UAVs choose the optimal resource allocation strategies depending on the network states autonomously. It was shown that the proposed approach yields up to $33.3\%$ and $50.3\%$ gains in terms of the number of users that have stable queues compared to Q-learning with cache and Q-learning without cache. It was also shown that LSM significantly improves the convergence time of up to $33.3\%$ compared to Q-learning.


\subsection{Ambient Backscatter Networks}
  
To increase SE, a cutting-edge technology named AmBC has received significant attention as a new SS framework \cite{amb0}. In backscatter communication (e.g., RFID), a device communicates by modulating its reflections of an incident RF signal without generating its own radio waves. Hence, it is in the orders of magnitude more energy-efficient than conventional radio communication.
AmBC system  enables two devices to communicate using ambient RF as the only source of power. It leverages existing TV and cellular transmissions to eliminate the need for wires and batteries, thus enabling ubiquitous communication where devices can communicate among themselves at unprecedented scales and in locations that were previously inaccessible.

\begin{figure}[h]
	\centering
	\includegraphics[width=2.7in]{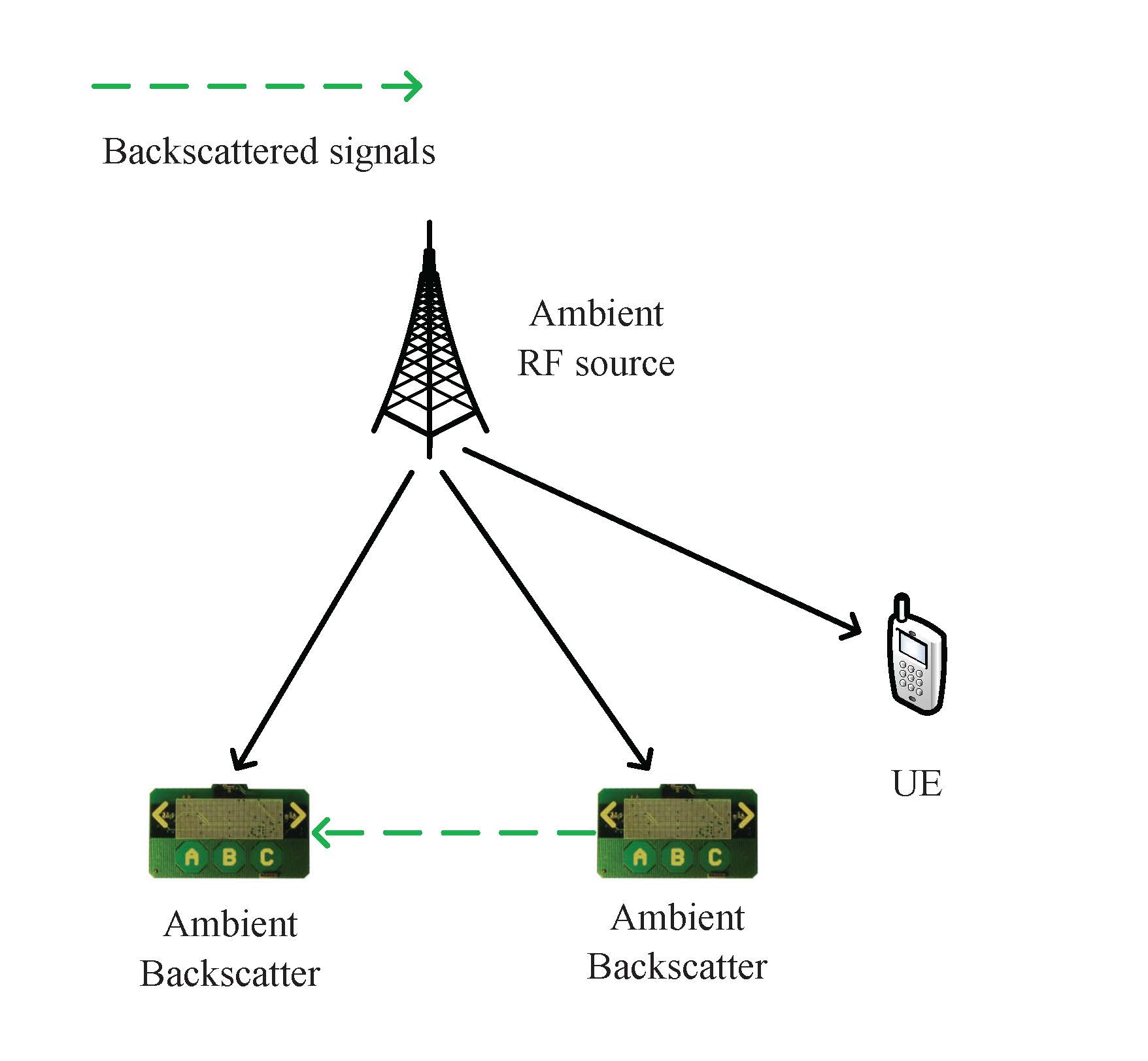}	
	\caption{ {AmBC network.\label{ambc}}}
\end{figure}

{In particular, in an AmBC system as illustrated in Fig. \ref{ambc}, the backscatter transmitter can transmit data to the backscatter receiver by modulating and reflecting surrounding ambient signals.} Hence, the communication in the AmBC system does not require dedicated frequency spectrum. Based on the received signals from the backscatter transmitter and the RF source or carrier emitter, the receiver then can decode and obtain useful information from the transmitter. 
By separating the carrier emitter and the backscatter receiver, the number of RF components is minimized at backscatter devices and the devices can operate actively, i.e., backscatter transmitters can transmit data without initiation from receivers when it harvests sufficient energy from the RF source \cite{amb01}. Therefore, AmBC systems can share spectrum with existing systems and achieve better spectral efficiency than that of RFID systems.

The existing ML based works for AmBC systems are mainly focused on the information extraction and mode selections.

\subsubsection{{Information Extraction}}

Since ambient backscatter uses uncontrollable RF signals that already have information encoded in them, it needs a different mechanism to extract the backscattered information. Several existing works has proposed different ML based methods to help extract information.


To solve the individual channel estimation in the AmBC system,  the authors in \cite{amb09} designed a communication protocol for the reader and the tag to obtain all the channel parameters. Based on the protocol, they proposed an ML-aided semi-blind estimator which utilizes an Expectation-Maximization (EM) algorithm and a few pilots from RF resources together with some superimposed pilots at the reader. The maximum likelihood estimator was applied to obtain the uplink channel between the reader and the tags while the superimposed pilots from the reader were used to estimate the downlink channel. Finally, the Cramer-Rao Bounds (CRB) of the proposed channel estimators were derived.

In an AmBC system, the readers receive the backscattered signal from the backscatter device (BD) and the Direct-Link Interference (DLI) from the RF source simultaneously. Due to the randomness of ambient RF sources, it is challenging to distinguish backscatter symbols from DLI. Furthermore, the existence of DLI  can further cause the conventional Energy Detector (ED) to fall into severe error floor problems. To tackle this issue, the authors in \cite{amb15} developed a novel error-floor free detector by using multiple receive antennas at the reader side. {They first considered the perfect CSI case and used beamforming-assisted ED and likelihood ratio based detector to decode the backscatter symbol.} Based on this, a novel statistical clustering framework was designed for joint CSI feature learning and backscatter symbol detection. It was verified that their method can achieve comparable performance with perfect CSI and significantly outperformed the conventional ED.


Recovering the tag information from the reader in an AmBC system is a challenging task when it is difficult to acquire the relevant CSI. To overcome this limitation, a label-assisted transmission framework was proposed in \cite{amb03}, in which two known labels were transmitted from the tag before data transmission. By exploring the received signal constellation information, they proposed a modulation-constrained expectation-maximization algorithm and two detection methods. The first method, named constellation learning with labeled signals, performs clustering only based on { the labeled signals and then recovers the unlabeled signals by using learned parameters.} The second method, named constellation learning with labeled and unlabeled signals, uses all received signals for clustering and can achieve better performance with higher complexity.

An ML-assisted AmBC information extraction method was proposed in \cite{amb06}. The information was modulated on top of the unknown Gaussian-distributed ambient RF signals. The binary phase-shift keying backscatter signals encoded by Hadamard codes can be decoded by the proposed method. By eliminating the direct path signal and correlating the residual signal with the coarse estimate of the ambient signal, the proposed method first extracted the learnable features for the tag signal. \emph{k}-nearest neighbors' classification algorithm was then employed to recover the tag signals. Finally, a Hadamard decoder was used to retrieve the original information bits from the recovered signals.


The energy detector or Minimum Mean Square Error (MMSE) detector utilized in existing AmBC systems to detect tag signals suffers from a high Bit Error Rate (BER). To overcome this challenge, Support Vector Machine (SVM) and random forest methods were proposed in \cite{amb05} for detecting the tag signals in an AmBC system by changing the detection problem into a classification problem. To minimize the BER, the proposed method could classified the received signals into different groups based on their energy features.


\subsubsection{{Operating Mode Selection and User Coordination}}

Due to its passive nature, Backscatter Devices (BDs) in AmBC systems must harvest energy to power operations such as circuit power consumption, transmission, and sensing. {BDs need to determine when to switch between communication and energy harvesting modes but the highly dynamic nature and randomness of RF source activities make this switch operation challenging \cite{amb07}.} Moreover, although the BD can perform the backscatter and energy harvesting simultaneously, it is impractical and inefficient when the amount of harvested energy is relatively small and can only supply internal operations. Therefore, how to efficiently determine the tradeoff between energy harvesting and backscattering RF signals is critical in a dynamic environment \cite{amb08}. Researchers have proposed various solutions based on RL algorithms.

By adaptively selecting the operating mode in a fading channel environment, the throughput maximization problem of the AmBC system was solved in \cite{amb07}. {With the given channel distributions, the problem was modeled as an infinite-horizon Markov Decision Process (MDP) and the optimal mode switching policy was obtained by the value iteration algorithm.} When the channel distribution information was unavailable, the Q-learning algorithm was employed to explore a suboptimal strategy through repeated interactions with the environment. {The efficacy of their proposed  Q-learning method showed that close-to-optimal throughput performance could be achieved.}


The authors in \cite{amb08} proposed an MDP framework to determine the optimal policy for allowing the secondary transmitter to maximize system throughput. The MDP-based optimization requires complete knowledge of environmental parameters such as the probability of a channel state and the successful packet transmission ratio. To cope with these impractical constraints, a low-complexity online RL algorithm was developed that allowed the secondary transmitter to learn from its decisions to discover the optimal policy. 
To minimize interference, a multicluster AmBC power allocation problem was developed in \cite{amb10} for short-range information sharing. A Q-learning-based power allocation method was designed to minimize the interference while improving the received SINR. It was shown that the received signal levels could be significantly improved by their proposed scheme.

By considering the strict latency requirements, the authors in  \cite{amb11} employed DQN to solve the communication rate maximization problem for wireless powered ambient backscatter tags. A Q-learning model for ambient backscatter scenarios was developed first, and an algorithm was then proposed that used DNNs to approximate the complex Q-network table. 



\subsubsection{{AmBC-CR Methods}}

Several works combined AmBC with CRN. An RF-powered backscatter CRN enables the secondary transmitter not only to harvest energy from primary signals, but also to backscatter these signals to the secondary receiver for data transmission \cite{amb12}. 
Such a combination can provide SUs with potential connection options instead of simply waiting for access opportunities.
When the primary channels in RF-powered CRNs that employ AmBC are mostly busy, instead restricting their activity to harvesting energy, the secondary transmitters can use a fraction of the wait time to transmit data by modulating and backscattering the received signals through the AmBC
Thus, AmBC enables secondary systems to maximize their performance by simultaneously optimize spectrum usage and energy harvesting.


In an AmBC-assisted CRN network, the mode selection between signal backscatter and energy harvest is critical to achieving high RF-powered SU (RSU) throughput. The dynamics of the primary channel, energy storage capability, and data to be sent all need to be considered when making decision.
An MDP-based framework was developed in \cite{amb13} to determine optimal decisions with consideration to states such as energy, data, and primary channel. It was then expanded to include a scenario in which the state information was unavailable at the RSU. A low complexity online RL algorithm was proposed to enable the RSU to find the optimal solution without requiring prior and complete information from the environment. 

In \cite{amb14}, the authors considered an AmBC enabled CRN, where multiple SUs communicate with a secondary gateway and operate in three modes: energy harvesting, backscattering, and actively transmitting. The gateway coordinates SUs by scheduling their backscattering time, energy harvesting time, and transmission time to maximize total throughput based on network states. The Double Deep-Q Network (DDQN) algorithm was employed to derive an optimal time scheduling policy under the dynamics of the primary channel and the uncertainty of the energy state of the SUs. 

\section{Security Issues in SS Systems}

\begin{figure}[h]
	\centering
	\includegraphics[width=3.0in]{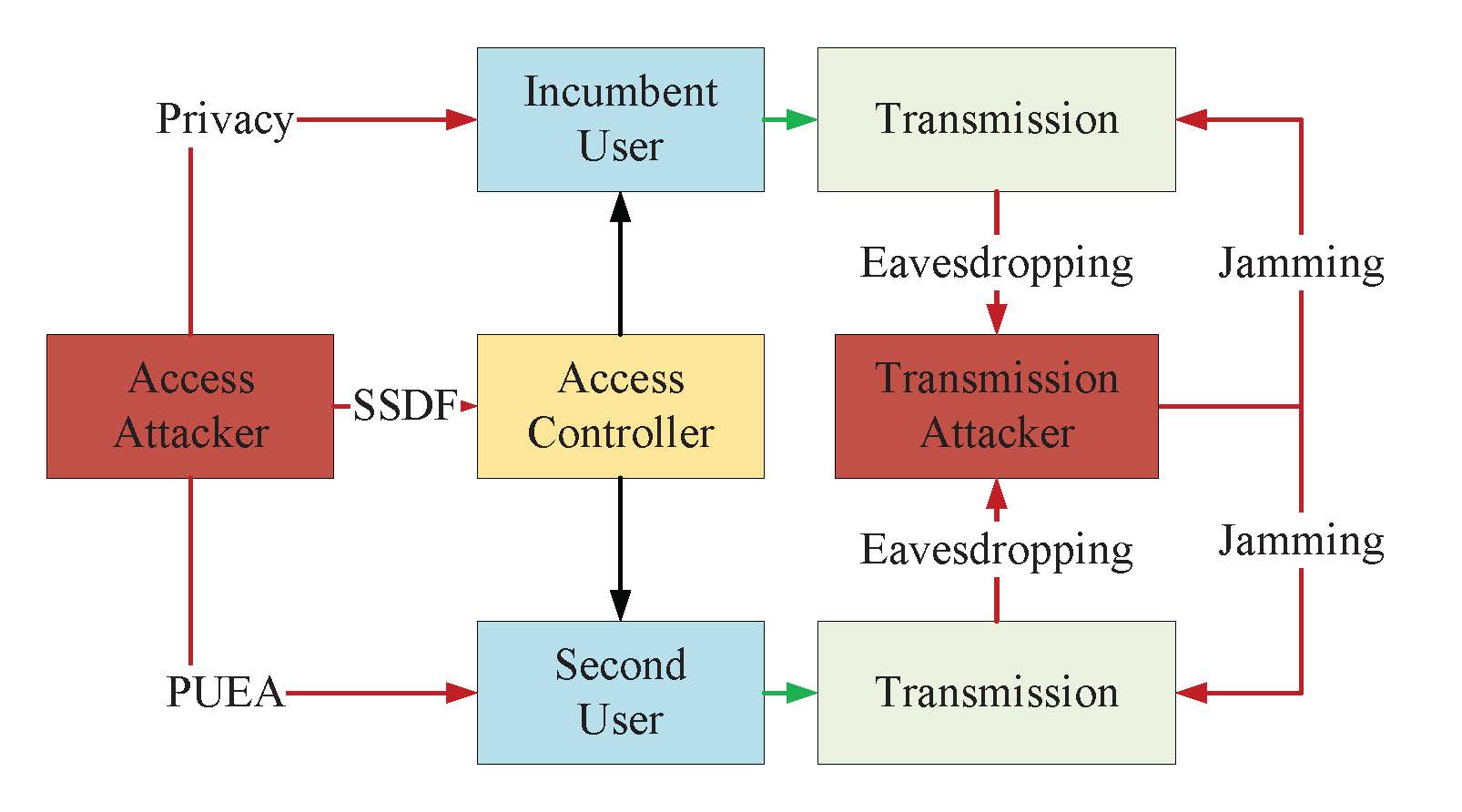}	
	\caption{ {Secure issues in SS network.\label{Sciss}}}
\end{figure}

While the ML-based SS networks can help improve the SE, they can also be a double-edged sword. The dynamic access frameworks introduce more security and privacy risks into the system. {As shown in Fig. \ref{Sciss}, when SUs observe the activity of PUs, the sensing procedure can be disturbed by the malicious attackers by launching the PUE attacks or SSDF attacks. The attackers may also exploit these opportunities to harm the privacy of PUs.} Besides these, the system also suffers the same security issues found in traditional wireless communications, such as jamming attacks and eavesdropping attacks. In this section, we will discuss these physical layer attacks and potential countermeasures.

\subsection{Primary User Emulation Attack} 

In CRN, a PUE attack denotes a PU-like signals sent by an attacker during the spectrum sensing period that can exclude legitimate SU access to the channels. The attackers may be selfish users who want to use the spectrum exclusively or malicious attackers who want to disrupt the normal operation of the system.
{PUE attacks can cause service degradation, denial of service (DoS), connection unreliability, and bandwidth waste \cite{puedc0}. }

{PUE attacks can damage required security such as availability, authentication, non-reputation, compliance, and access control \cite{sssi01}.} Countermeasures to PUE attacks seek to enhance spectrum management. When defending against PUE attacks, it is important to differentiate between malicious users and legitimate users. This can be determined by their location, received signal strength, received signal power, and other features.  

\subsection{Spectrum Sensing Data Falsification Attack}
{CSS as one promising approach for PUs' activities detection involves exploiting the spatial location diversity of multiple SUs.}
A group of SUs collaborate to perform the spectrum sensing by exchanging locally-collected information. An SSDF attack (also known as the Byzantine attack) is launched in CSS by sending false local spectrum sensing results to others, leading to flawed spectrum sensing decisions \cite{ssdf0}.
{SSDF attacks aim to decrease detection probability and disturb normal operations of the primary system.} It may also seek to increase the probability of false alarms in order to deprive honest SUs of access opportunities \cite{ssdf00}. SSDF attacks harm the system's integrity and availability.

 SSDF attackers can be classified into three types: selfish SSDF, interference SSDF, and confusing SSDF \cite{ssdf0}. 
 \begin{itemize}
     \item[(1)] A selfish SSDF attacker seeks to gain exclusive access to the target spectrum. It falsely reports the existence of relatively high PU activities to block other SUs from using the spectrum when the PU does not exist.
\item[(2)] {An interference SSDF attacker falsely reports low PU activities leading other SUs to wrongly conclude that they can use the spectrum without interfering with any PUs.} This type of attack seeks to either cause the inference to the PU or inhibit the communication of other SUs.  
\item[(3)] A confusing SSDF attacker seeks to disturb the SUs to prevent them from reaching consensus by randomly reporting the true or false results about the existence of PUs.
 \end{itemize}

The majority of existing defense methods can be divided into two categories: 
one making direct judgments based on the current spectrum sensing data while the other uses the historical spectrum sensing data to update sensors' reputation.


 \subsection{Jamming Attacks}
The open nature of wireless communication leaves it vulnerable to various attacks. 
One of the most common attacks in wireless communication as well as SS networks is the jamming attack. Attackers transmit signals to interfere with the victims' communications in order to cause a DoS and compromise availability of communication links \cite{jam00}. Traditional anti-jamming methods used in wireless communications include sequence-based frequency hopping spread spectrum (FHSS) and direct sequence spread spectrum (DSSS). However, the fixed transmission patterns of these methods leave them helpless against dynamic jamming attacks and cause low spectrum efficiency.  

SS techniques enable flexible access to different channels, allowing users to avoid attackers by exploiting that flexibility. 
The ML techniques provide more adaptive channel selection ability to systems in order to avoid jamming attacks. They also give the system the ability to learn and predict the behavior of jammers to increase anti-jamming channel selection efficiency.
The attackers may also use different ML based methods to improve their attack strategies \cite{jam02} rendering the study of advanced jamming attacks and corresponding countermeasures of vital importance to the SS system.
 
\subsection{Intercept/Eavesdrop}
Eavesdropping is another common attack in wireless communications. Due to the broadcast nature of radio propagation, any active transmissions operated over the shared spectrum by different wireless networks are extremely vulnerable to eavesdropping. It is therefore important to investigate the confidentiality protection of SS communications against eavesdropping attacks \cite{eve00}.

There are two major categories of secure communication techniques that guard against eavesdropping. One focuses on traditional cryptographic techniques and the other is the physical layer security. Cryptographic techniques involve encryption and decryption of information at the transmitter and receiver.
In the physical layer security method, the secrecy rate can be achieved by the mutual information difference between the legitimate user and the eavesdropper. However, the security rate can be limited since it depends on the difference between the channel condition from the transmitter to the legitimate receiver and that from the transmitter to the eavesdroppers. Many promising techniques have been proposed to address this issue, including artificial noise (AN) and cooperative jammer (CJ) \cite{qwirs}. The advantage of physical layer security over cryptographic is that it can achieve secure communications without extra overhead caused by protecting the security key and can therefore be used in relatively simple communication systems.

\subsection{Privacy Issues in Database Assisted SS Systems}
{According to \cite{prcy001}, there are several differences between security issues and privacy issues.}
Security issues refer to unauthorized/malicious access, change, or denial of data. Privacy issues refer to the unintentional disclosure of sensitive information from some open-access data. {The former is usually the work of malicious attackers who wish to disturb the system. In the latter, malicious users usually only collect information that does not immediately cause direct harm to the system.} The goals of security protection are confidentiality, integrity, and availability. The goals of privacy protection are anonymity, unlinkability, and unobservability.


Ensuring privacy in SS networks is very important. In some SS systems like SAS in the CBRS band, the IUs can be radar devices and military ships carrying very sensitive information. Opening these incumbents' exclusive spectrum to sharing could usher in potential privacy threats to the system. Furthermore, the distribution structure of the SS network increases the risk of privacy leakage. Finally, in order to train itself, ML requires huge amounts of data that may contain various private user information that must be protected during training and communication.

One possible attack is the database inference attack (DIA), where malicious users can obtain PU location and other private information through collected data and sophisticated inference techniques. In another form of attack, the operational privacy threat of SUs, which comes from the untrustworthy database that collects the location information sent from SUs on the set of available channels in their region.

The protection of PUs privacy cannot be addressed by strictly controlling access to the database, since each SU must access it to enable the spectrum sharing process.
One possible solution might be to reveal obfuscated information instead of the original information to SU queries. By doing this, the system can use the obfuscated information to help determine the channel status while reducing leakage of PU's privacy information.
{SUs' privacy can also be protected by sending an obfuscated version of the original information of SUs.} Some works also looked at the question of how much information to share with the database and the dynamic question of whether to share information with the database.

ML algorithms require massive amount of data to train their models. These data usually include a lot of user-specific sensitive info and need to be exchanged in some distributed systems. Sensitive information may leak out during the training process that would have remained secure using the above spectrum sensing procedure.
Three main strategies may be used to maintain privacy in ML work flow: differential privacy, homomorphic encryption and Secure Function Evaluation (SFE)/Secure Multi-party Computation (SMC) \cite{prcy01}.
In the differential privacy method, publicly shared dataset information describes the patterns of groups within the dataset but withholds information about individuals.
In homomorphic encryption, the operation on encrypted data can be used to secure the learning process by computing on encrypted data.
 When user-generated data are distributed among different data owners, SFE can enable multiple parties to collaboratively compute an agreed-upon function without leaking input information regarding any party other than what can be inferred from the output.




\section{ML-assisted Secure SS}
\begin{figure}[h]
	\centering
	\includegraphics[width=3.3in]{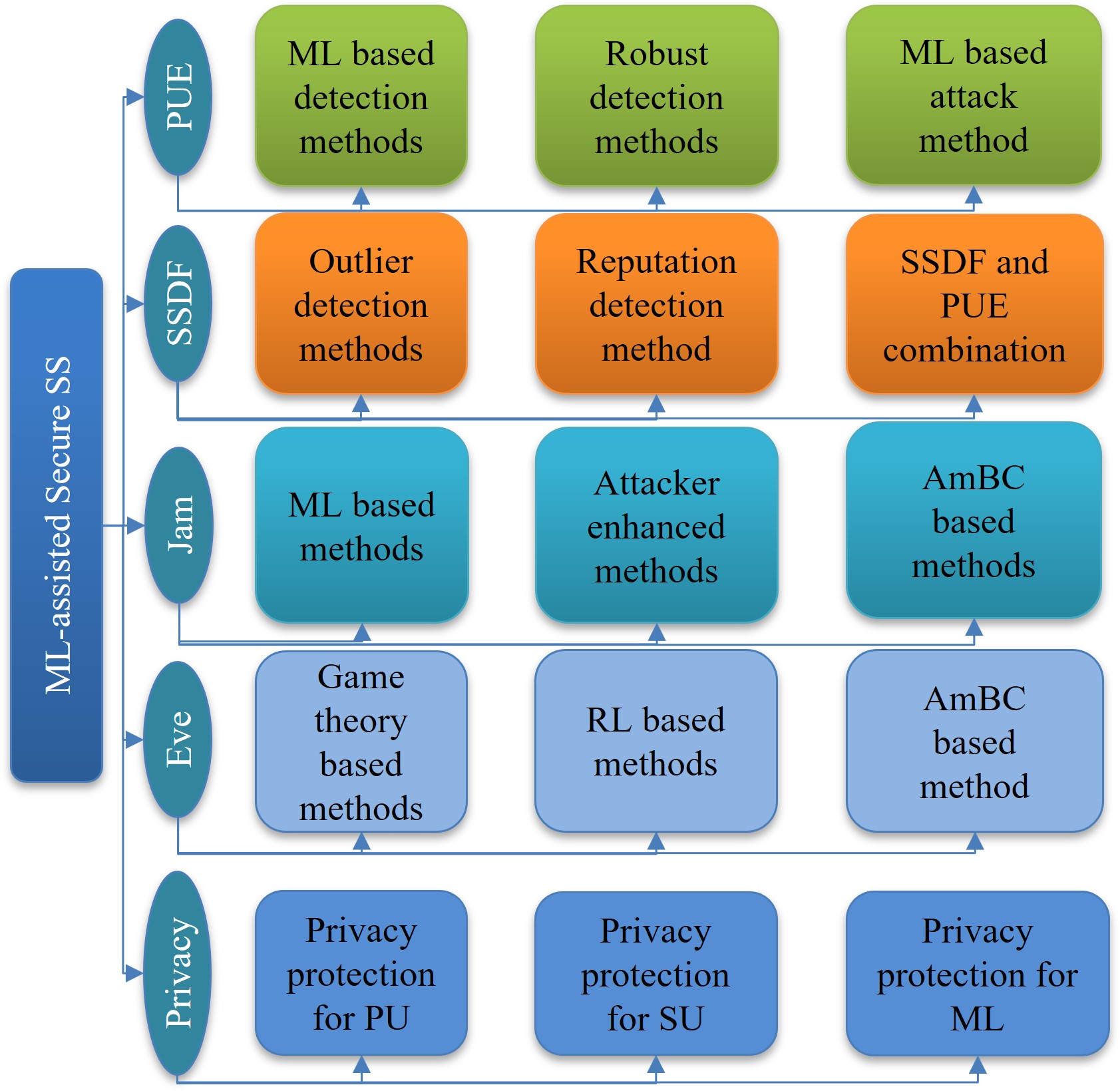}	
	\caption{ {ML-assisted Secure SS.\label{secure1}}}
\end{figure}

In this section, the latest research in ML related security will be comprehensively surveyed. The contents are organized as in Fig. \ref{secure1}. Details of existing works will be reviewed in each category.

\subsection{State-of-the-Art Methods of Defense Against PUE Attack}
 In this subsection, different ML-based PUE attacker detection methods in \cite{puedc9, puedc8, puedc1, puedc11, puedc10, puedc14} will first be presented. To address the limitations of training data and inconsistent communication environments, the robust detection methods in\cite{puedc3, puedc4, puedc2, puedc7} will be further discussed. Finally, the ML-based attack strategies included in \cite{puedc13, puedc5, puedc6} will be discussed to provide an attacker's perspective on PUE defense method design.

\subsubsection{ML Based Detection Methods}

\begin{figure}[h]
	\centering
	\includegraphics[width=1.8in]{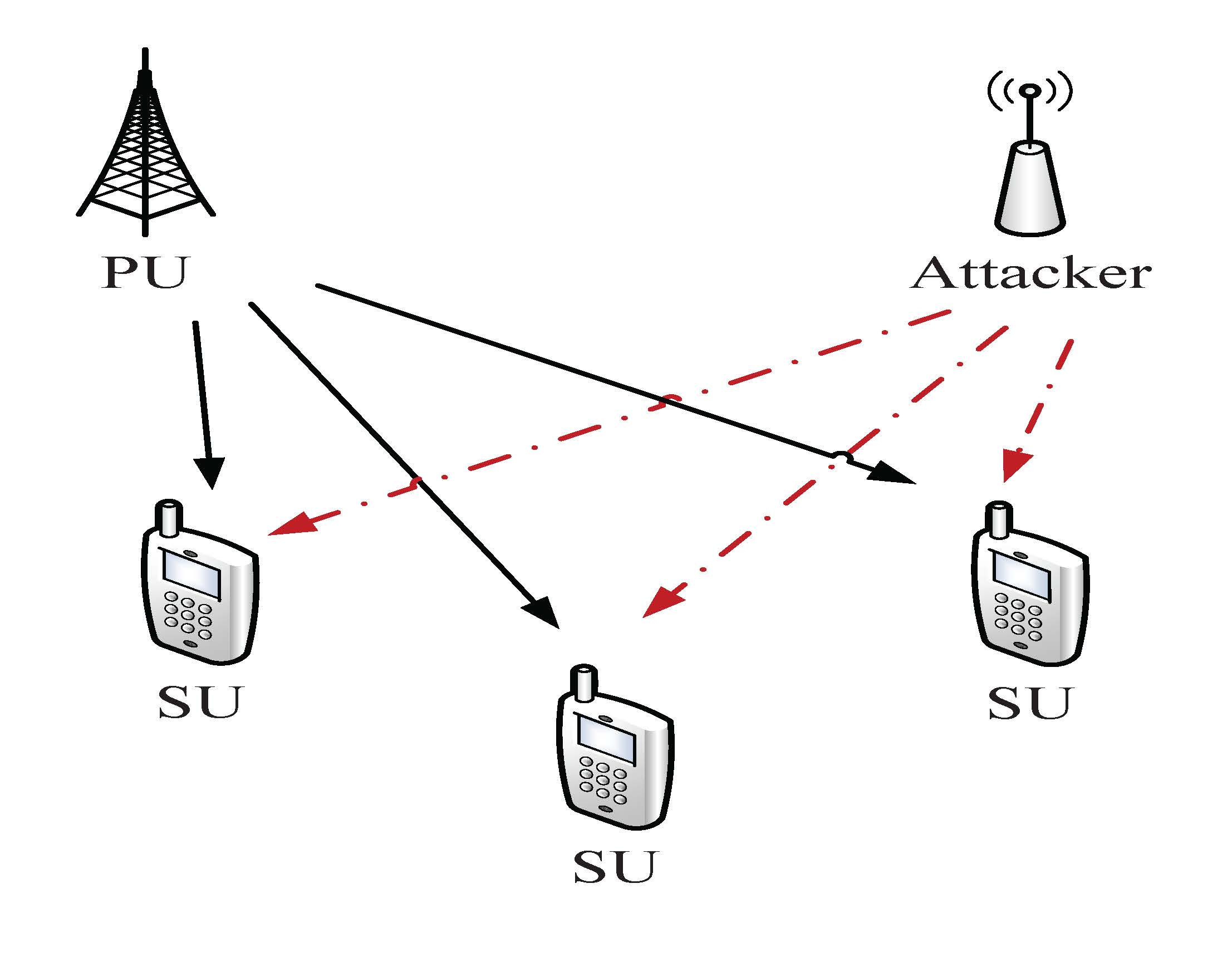}	
	\caption{ {Illustration of PUE attacks.\label{puea}}}
\end{figure}

A typical PUE attack is illustrated in Fig. \ref{puea}. In defending against such attacks, the most important step is distinguishing malicious attackers from legitimate PUs. This can be achieved using specific features extracted from received signals. Distinct features may reflect the transmitters' characters, rendering them unique and differentiable. {User location based method is a common and easy way to differentiate between attackers and PUs.} Malicious attackers are rarely in the same place as PUs. Since the received signal strength (RSS) varies by location, it can be adopted to identify location and, by the same token, user type. Some other methods are based on statistical analysis. They use features such as signal power, spectrum occupancy time, and cyclostationarity extracted from received signals to analyze transmitters. Finally, the physical layer approaches uses the hardware behaviors of transmitters or channel behaviors to detect attackers. For example, phase and frequency shifts are commonly used as transmitter fingerprints. 
A detection problem based on received signals is a classification problem, which ML is particularly good at solving. 

To detect and defend against PUE attacks, an adaptive learning-based detection method in CRN was developed in \cite{puedc9} by analyzing the transmitters’ power features. Specifically, cyclostationary feature analysis was used to differentiate attackers from low-power PUs. The proposed method also estimated the distance variance and communication time to improve classification accuracy and communication rate.

A k-nearest neighbor (KNN) classifier based detection method was used to classify malicious users to forestall PUE attacks in \cite{puedc8}.
{KNN was trained with the parameters such as data rate, distance, power, frequency of request.} Moreover, Elliptical Curve Cryptography (ECC) was applied to encrypt the data and improve network security. 
{The proposed classifier achieved a higher accuracy detection performance than the Artificial Neural Network (ANN) based method.}

A channel-based method that relied on the behavior of the multi-path channel  was investigated in \cite{puedc1}, where the authors proposed an ML framework based on various classification models for detecting PUE attacks. It was trained/tested using four features vectors extracted by the Pattern Described Link-Signature (PDLS) method. 
By using this method, legitimate and malicious users could be effectively distinguished. 

When signal activity patterns can be regarded as a possible sequence relating to some “features” of the channel and previous internal states, the Recurrent Neural Network (RNN) is a good tool for PUE detection.
An RNN based PUE attack detection method was first introduced in \cite{puedc11}. It exploited series' temporal dependency for better series prediction and abnormal activity detection. {To deal with the gradient vanishing issues inherent to RNN, an advanced version of RNN that took advantage of the Long Short Term Memory (LSTM) units and processed time series with long-term memory more efficiently was further proposed.} It was shown that the LSTM based method could significantly improve the detector’s performance.

The authors in \cite{puedc10} investigated the joint detection of PUE and jamming attacks in CRN. A sparse coding of the compressed received signal based detection algorithm was proposed. Based on the channel dependent dictionary, convergence patterns in sparse coding were employed to differentiate the spectrum hole, legitimate PU, and emulators or jammers. An ML based classification was adopted to perform the decision making operation. The effectiveness and advantages  of the proposed algorithm were verified in terms of the confusion matrix quality metric.

 By detecting PUE attacks and enhancing the probability of detection, a hybrid Genetic Artificial Bee Colony (GABC) algorithm was proposed in \cite{puedc14} to optimize spectrum utilization. 
 A Genetic Algorithm was used to compensate for the Artificial Bee Colony algorithm’s less than optimal exploitation of solutions by using crossover and mutation operations.
 The proposed GABC incorporated the Genetic operators into the Artificial Bee Colony algorithm to achieve balance between exploitation and exploration in order to find the optimal solution.

\subsubsection{Robust Detection Methods}
Most of the classification approaches discussed above require a certain dataset to train the network. However, an unfamiliar environment and attackers may lead to classifier failure due to an unsuitable reference model. Hence, robust detection methods are very important when adapting to the changes in the environment and more practicable in the real world.

In \cite{puedc3}, defense strategies against PUE attacks from malicious SUs were investigated. Greedy SUs manipulated unsupervised learning clustering algorithms through various attack strategies to evict other SUs from the idle channel. Corresponding countermeasures to these manipulation attacks were developed, and the robustness of unsupervised learning for signal classification in an evolving RF environment was evaluated. By using both $k$-means clustering and Self-Organizing Maps (SOMs), the proposed method could perform signal classification in the absence of training data. With intentional attacks from SUs, the robustness of the classifier to avoid misclassification was verified. It was shown that the efficacy of attacks could be reduced by $75.9\%$.

By adopting the TL algorithms, the authors in \cite{puedc4} developed a PUE defense approach that used knowledge about PUs and SUs from past time frames to improve the detection process in future time frames. The proposed approach extracted high-level representations of the environment and accumulated them to form an abstract knowledge database. This database enabled the CR system to accurately detect PUE attacks even if an insufficient amount of fingerprint data was available in the current time frame. The final detection decisions were used to update the abstract knowledge database for future runs.

A semi-supervised distributed learning algorithm was proposed for PUE attack detection. By enabling edge devices to perform data clustering and session classification locally, it could deal efficiently with varying bandwidth, signature changes, etc. The labeled data was feed into a trained supervised learning based classifier for classification. Based on the error rate, it adjusted the training vectors and improved the overall performance. It was shown that the proposed method could significantly reduce false alarms in the secondary network and improve overall detection accuracy in the primary network.

In \cite{puedc7}, an adaptive Bayesian learning automaton algorithm-based scheme named Multi-channel Bayesian Learning Automata (MBLA) was proposed to defend against PUE attackers. The SU in the considered model adopted Uncoordinated Frequency Hopping (UFH) to avoid PUE attacks. To improve the speed and accuracy of the learning process in non-stationary environments, MBLA utilized two different channels simultaneously to perform the optimal frequency channel selections. Statistical information about channels and PUs was assumed unavailable. An SU synchronized with its receiver and sent its data on various channels obtained by the MBLA. The scheme extracted the best strategies for the attacker and the SU and then evaluated the proposed scheme in terms of the SU throughput in the presence of the PUE attacker.

\subsubsection{ML Based Attack Methods}
Throughout the above discussions, various PUE attack detection and defense strategies have been offered. However, the best-attack strategies have not been discussed. A better understanding of optimal attack strategies can enable researchers to quantify the severity or impact of a PUE attacker on an SU’s throughput. It can also shed light on practical defense strategy design as the attackers can also exploit  ML algorithms to improve their performance.

 In \cite{puedc13}, the authors presented two GAN based models that successfully emulated the PUs. Depending on whether any prior knowledge of the PU's feature space available, they proposed a dumb generator model and a smart generator model. {Two DNN based discriminator models were developed to distinguish the PU and the Emulated PU (EPU) from the corresponding generators.} With iterative and sequential training, the generator and discriminator of each GAN model became smarter and smarter. It was shown that discriminators were able to detect  about $50\%$ of PUE attackers without the GAN training during the deployment phase and both the GAN models could achieve $100\%$ accuracy during the training phase. After the GAN training, the discriminators of the dumb generator could achieve $98\%$ accuracy while the smart generator based model could achieve $99.5\%$ accuracy.


Optimal PUE attack strategies were investigated in \cite{puedc5}, where prior knowledge on PU activity characteristics and SU access strategies were not available. Based on previous attacking experience, a non-stochastic online learning problem was formulated to determine attacking channel decisions for attackers. Since a PUE attacker never knows if an SU has ever tried to access the attacked channel or not, it cannot observe the reward. Therefore, an Attack-But-Observe-Another (ABOA) scheme was proposed to solve this issue. The attackers in this scheme  attack one channel in the spectrum sensing phase but observe one or more other channels in the data transmission phase. Two non-stochastic online learning-based attacking algorithms, EXP3-DO and OPT-RO,  were proposed to select the observing channel deterministically based on the attacking channel and uniform randomly, respectively.
{The authors in \cite{puedc6} proposed an online learning-assisted attacking method based on the attacker’s observation capabilities.} Their work showed that the attacker loses on the regret order when there is no observation within the attacking slot. However, the attack performance can be significantly improved with the observation of at least one channel. {Moreover, observation of multiple channels cannot benefit the attacker more, though it gives insight into the number of observations required to achieve the minimum constant factor.}


\subsection{State-of-the-Art Methods of Defense Against SSDF Attack}
\begin{figure}[h]
	\centering
	\includegraphics[width=3.0in]{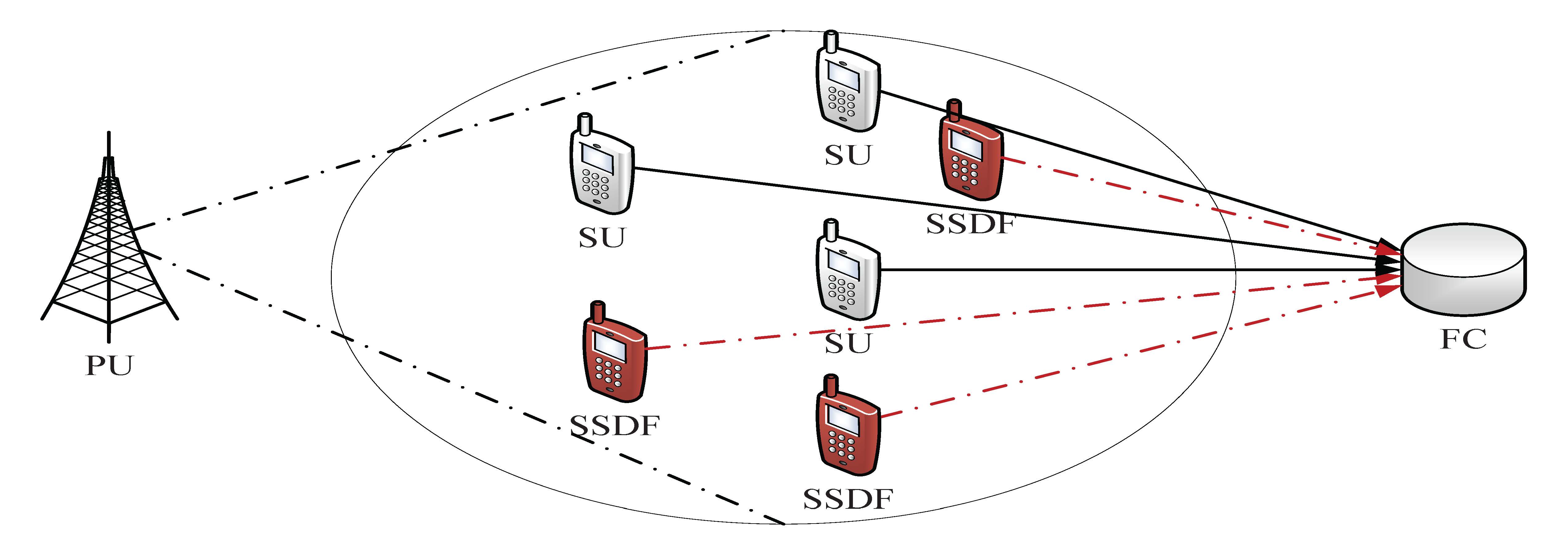}	
	\caption{ {Illustration of SSDF attack.\label{SSDF}}}
\end{figure}

CSS can help overcome the fading environments and improve the system sensing performance. Different from single-user-based SS, each SU needs to transmit the sensing results to a Fusion Center (FC) in CSS. FC then combines those results and makes a final decision about the PU's presence. SSDF is the most common attack in CSS. {As shown in Fig. \ref{SSDF}, sending falsified sensing data to the FC can lead to an incorrect fusion result, cause interference with PUs, and cause DoS to SUs.}
To defend against SSDF attacks, the most important step is to differentiate attackers from legal SUs. The existing defense methods fall into two groups, namely outlier detection approaches and reputation-based approaches. In outlier detection methods, the abnormal user is excluded from the network. In reputation-based methods, on the other hand, SUs are assigned a reputation degree that reflects their detection performance. Since  SUs are not eliminated and their reports are not excluded, reputation-based methods can use the collected information more thoroughly than outlier detection techniques. 


\subsubsection{Outlier Detection Methods}
To mitigate SSDF attacks, the authors in \cite{ssdf12} proposed a Support Vector Data Description (SVDD) algorithm in the sensing phase that could distinguish malicious nodes from legal ones and remove them in the decision phase. The boundary around the target data was constructed by enclosing the target data within a minimum hyper-sphere. Enlightened by the SVM, the SVDD decision boundary was described by a few support vectors. {The spectrum sensing result was then decided according to the voting results from trusted nodes. } 

 In \cite{ssdf5}, the authors designed a Bayesian nonparametric clustering scheme to sense the primary channel status and identify malicious users in CSS. By forming a single cluster consisting of all legal users and a separate cluster that included every variety of attacker (selfish, mischievous, jamming) in the feature space, their proposed approach could detect and identify multiple attacks simultaneously. Furthermore, based on observations from the collaborating CR users, it could discern malicious users from the legitimate ones and obtain the real PU traffic pattern.

In \cite{ssdf8}, Bayesian learning-based SSDF defense schemes were proposed. First, a Bayesian offline learning algorithm was proposed where the spectrum state was unavailable for training. A Bayesian online learning framework was then designed by incorporating the time-varying attributes of the sensors. The framework consisted of the historical data learning part and the current data learning part. The vector of sensors’ weight was updated by considering both the historical and the current spectrum sensing data. {Finally, an SSDF attack behavior recognition algorithm based on the proposed framework was designed to identify SSDF attacks more accurately than offline learning.}


\subsubsection{Reputation Based Detection Methods}

One of the critical issues in combating SSDF attacks is distinguishing the attackers' error reports from the SUs' reports in the FC. A Bayesian reputation model based SSDF defense scheme for CRNs was proposed in \cite{ssdf4}. The proposed method modeled cooperation as a service-evaluation process and SUs' reputation degrees reflected their service quality. Reputation degrees of SUs were updated based on the Bayesian reputation model, and the behaviors of malicious SUs could be effectively tracked by this means.

In \cite{ssdf6}, a three-layer Bayesian model was designed to combat SSDF attacks. The model consisted of a processing layer, an integrating layer, and an inferring layer. The processing layer was based on the HMM model, where original data was used to train parameters and the trained emission distributions were then passed to the second layer. By employing different algorithms in the integrating layer, emission distributions were processed to obtain reputation values, balance values, and specificity values of different SUs. By using different thresholds, these continuous values could be rendered discrete and then transferred to the inferring layer. Finally, by using the discrete values as evidence, a Bayesian network was built in the third layer to calculate the safety probabilities of SUs.

In \cite{ssdf16}, several ML techniques such as SVM, Neural Network, Naive Bayes, and Ensemble classifiers were implemented to detect SSDF attacks in a CRN. The learning techniques were investigated under two experimental scenarios: (a) the training and test data were drawn from the same data-set, and (b) separate datasets were used for training and testing. The robustness of the proposed ensemble method was verified compared with other benchmarks.

An SVM-based scheme was proposed in \cite{ssdf13} to deal with SSDF attacks. SUs' behaviors were analyzed from multi-round records of energy values, and their classification accuracy was obtained. Furthermore, the concepts of recognition probability and misclassification probability were introduced, and the tradeoff relationship between misclassification probability and threshold of classification accuracy were obtained. {As a result, the proposed scheme enabled excellent adaptability for Malicious SU (MSU) detection in various scenarios.}

A distributed CSS method based on RL was proposed in \cite{ssdf18} to remove data fusion between different reputations-based users in CRN. SUs acted as agents which were selected from the adjacent nodes of CRN participating in the CSS. The reputation value was employed as the reward function to ensure that the agents merged with high reputation nodes. To improve the consensus of the whole network, conformance fusion was adopted and compared with the decision threshold to complete the CSS. It was shown that the proposed method could identify the attackers and achieve stable performance.

\subsubsection{SSDF and PUE Combination Attacks}
The combination of the SSDF attack and the PUE attack presents more challenges to the network. If SUs are be attacked or mislead by the PUE attack, the performance of existing SSDF defense methods is degraded. Even if the PUE attacker is detected, neighboring SUs can still submit flawed sensing reports due to the contaminated signal from the PUE attack. Investigating the combination of attacks and corresponding ML-based countermeasures in existing works can provide a comprehensive understanding of secure design in SS networks.

Secure sensing under both PUE and SSDF attacks in CRN was investigated in \cite{ssdf22}. Directly excluding attacked SUs from the sensing cooperation process requires lots of information, i.e., the attack strength, geographical locations of SUs, etc. Therefore, a novel secure sensing algorithm was developed to deal with the problem. To be specific, Unsupervised ML (UML) was adopted to identify contaminated sensing reports from trusted users by examining their sensing history. These contaminated sensing reports were then excluded from CSS. Moreover, considering that identification errors might occur during the UML process, each SU was assigned an identity value to account for its reliability. The identity value was also used to alleviate the misidentification impact on real trusted users.

To defend against various malicious attacks and interference in full-duplex CRNs (FD-CRNs), an ensemble ML (EML) based robust CSS framework was proposed in  \cite{ssdf24,ssdf25}. SUs were assumed to have the ability to sense and transmit over the same frequency band simultaneously. The self-interference and co-channel interference were inevitably introduced into the system and complicated the sensing environment. By investigating spectrum waste probability, collision probability, and secondary throughput in both FD LBT and Listen-and-talk protocols, the robust and accurate fusion performance of the proposed EML approach was verified.

 \subsection{State-of-the-Art Methods of Defense Against Jamming attacks}
 
{A jamming attack is a common attack in wireless communication systems and there have been many works on this subject.}
 \subsubsection{ML Based Anti-Jamming Methods}
In the SS network, except for the very rapid change of dynamic spectrum characteristics in the channel, the inclusion of random jammers makes efficient communication more challenging.
To defend against jamming attacks, the most common countermeasure strategies are dynamic channel assignments based on different ML algorithms.
The jamming attack scenario can be modeled by using the stochastic zero-sum game and MDP framework. {The time-varying characteristics of the channel as well as the jammer's random strategy can be learned by the SU using RL algorithms.}

 A stochastic game framework was proposed in \cite{jam0101} for anti-jamming defense. At each stage of the game, SUs observe the spectrum availability, the channel quality, and the attackers' strategy from the status of jammed channels.
Based on observation results, the number of reserved channels and the channel switch action policy are decided. SUs employ Minimax-Q learning to learn the optimal policy, maximizing the expected sum of discounted spectrum-efficient throughput. {It was shown the proposed stationary policy in the anti-jamming game performed better than the myopic learning and random defense strategy because it successfully accommodated the environment dynamics and strategic behavior of the cognitive attackers.}
To further improve system performance, the authors used the QV and the SARSA RL algorithms in \cite{jam01} to replace the Minimax-Q learning in \cite{jam0101}. Minimax-Q learning is an off-policy and greedy algorithm, whereas the QV and SARSA are on-policy algorithms. It was shown that QV learning can achieve the best performance as the value of $Q$ and $V$ are both updated.

 
In \cite{jam02}, the authors first investigated an anti-jamming game model where the SU could access only one channel at a time and hopped among different channels. An MDP-based channel hopping defense strategy with the assumption of perfect knowledge was derived by analyzing interactions between the SU and attackers. Based on this, they proposed two learning schemes by which SUs gained knowledge of adversaries in order to handle cases without perfect knowledge. The schemes were then extended to a scenario where SUs could access all available channels simultaneously, and redefined the anti-jamming game with randomized power allocation as the defense strategy. The Nash equilibrium was derived for this Colonel Blotto game, which minimized the worst-case damage.

In \cite{jam03}, a game model was formed to integrate anti-jamming and jamming subgames into a stochastic framework. Q-learning was applied to find an optimal channel access strategy. It was shown that Minimax-Q learning was more suitable than Nash-Q learning for an aggressive environment. For distributed mobile ad hoc networking scenarios, Friend-or-foe Q-learning provided the best solution where centralized control was nearly unavailable.

{By employing the Double Q-learning algorithm to defend against the jamming attacks, Multi-Objective Ant Colony Optimization (MOACO) and greedy-based optimization methods were proposed in \cite{cr015}.} A Q-learning assisted cluster-based data utilization was proposed that could enhance inter-cluster data aggregation. The network lifetime was improved using AI-based modeling with intra-network to enhance green communication. Unlike the artificial bee colony and genetic algorithm, the throughput, device lifetime, and jamming prediction were promoted using the proposed MOACO.

A Wideband Autonomous CR (WACR) anti-jamming method presented in \cite{jam07} evaded a jammer that swept across the whole wideband spectrum range. The WACR equipped spectrum knowledge acquisition ability to detect and identify the location of the sweeping jammer. A Q-learning-based method was proposed to allow the anti-jamming operation to cover over several hundred MHz of a wide spectrum in real-time. An anti-jamming-based secure communication protocol was then developed that selected a spectrum position with enough contiguous idle spectrum to resist interference by both deliberate jammers and inadvertent disruptions. The communication then switched to this position until the jammer arrived. When the jammer began to interfere with the CR's transmission, it switched to a new spectrum band that lead to the longest possible uninterrupted transmission as learned through Q-learning. 
By including more agents in the system, the authors in \cite{jam08} further proposed an advanced RL-based anti-jamming approach. The considered system model allowed multiple WACRs to operate over the same spectrum band simultaneously. Each radio attempted to evade other WACRs' transmissions and avoid jammer signals that swept across the whole spectrum band of interest. The WACR first detected and identified the frequency location of this sweeping jammer and the signals of other WACRs. A sub-band selection policy was then given by the RL-based approach based on the detection results to avoid both the jammer signal and interference from other radios.

To enable network devices to detect and predict jamming signals in a system with multiple jamming modes and noises, it is critical to develop a rapid jamming detection countermeasure. To this end, the authors in \cite{jam018} proposed a DL-based jamming pattern recognition by using spectrum waterfall. In addition, the simplified Le-Net5 structure was employed to reduce the complexity of the calculation. As a result, the proposed method achieved a rapid recognition performance.

By directly using temporal and spectral information like spectrum waterfall, the authors in \cite{jam020} developed a novel anti-jamming approach that did not require knowledge of jamming patterns and parameters. First, a recursive CNN was designed to overcome the infinite-state issues of spectrum waterfall. Furthermore, a DRL algorithm-based anti-jamming method relying only on locally observed information was proposed to obtain optimal anti-jamming strategies. The proposed method could explore the unknown environment and combat advanced jamming attacks in a more practical fashion.

{A sequential DRL algorithm without prior information was proposed in \cite{jam019} to defend against jamming attacks.} The jamming patterns were first identified by DL and sliding window principles. Those recognized patterns were then passed to an RL based model to inform online channel selection. To better achieve the tradeoff between throughput and overhead, channel switching cost was introduced to the system. It was shown that the proposed method could make anti-jamming channel selection decisions quickly without modeling the jammer's characteristics.

\subsubsection{Attacker Enhanced Anti-Jamming Methods} 
 \begin{figure}[h]
	\centering
	\includegraphics[width=3.3in]{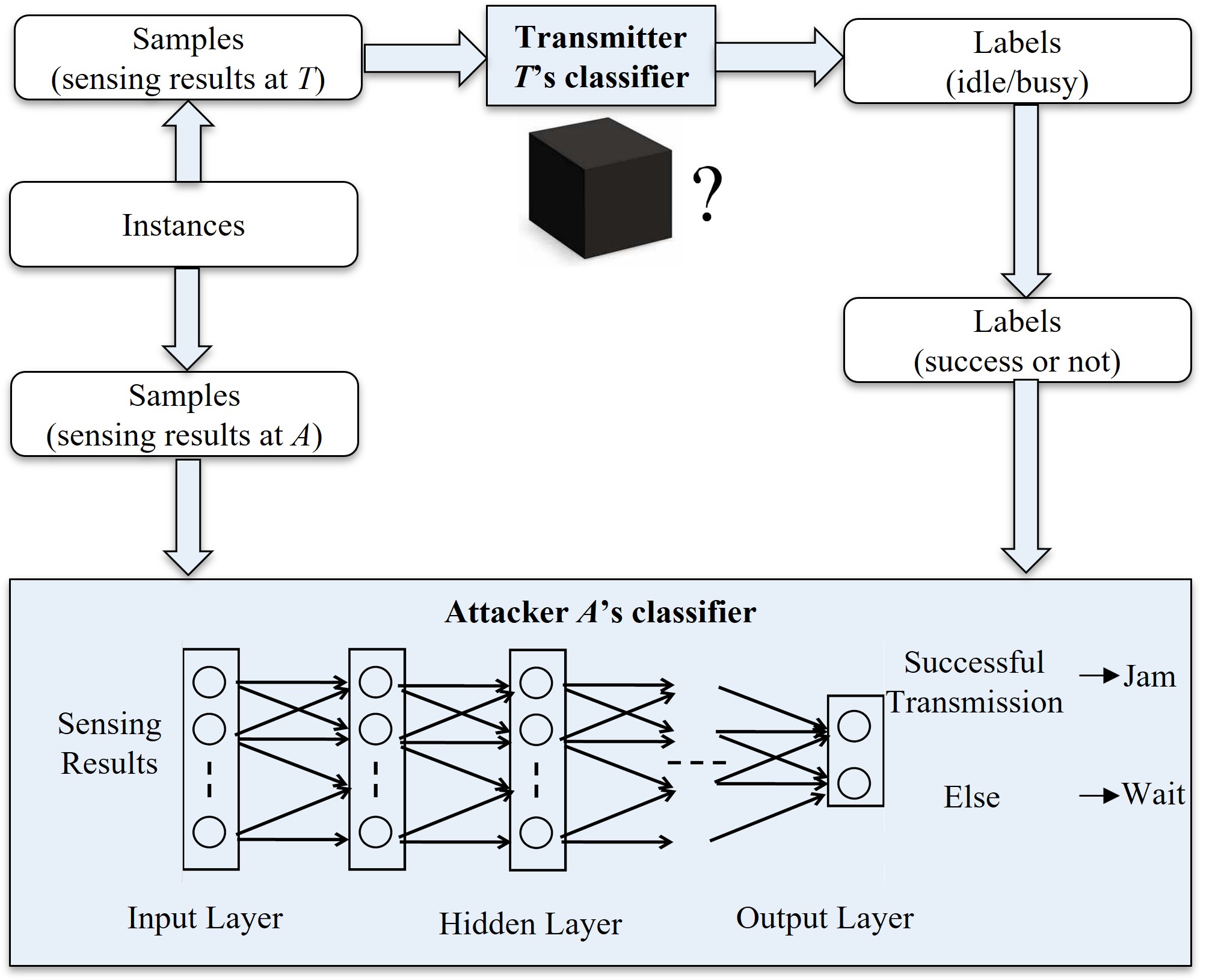}	
	\caption{ {System model for attacker’s learning \cite{jam012}.\label{Jam1}}}
\end{figure}
 
Although ML provides many effective solutions for system defense when fighting jamming attacks, it can also be exploited by attackers to develop more effective attack strategies. Considering intelligent attackers when designing defensive measures can help  avoid overly clumsy assumptions and enhance protection schemes more reliable and practical.


In addition to anti-jamming techniques, knowing intelligent jamming strategies is also crucial. An intelligent jammer that can adapt to its surroundings was investigated in \cite{jam04} under an electronic warfare type scenario. To be more practical, the delay of packet exchange information between the victim senders and the receivers was considered by the jammer, as opposed to the traditional assumption where the feedback is instantaneously available. Furthermore, to implement delayed learning in scenarios where rewards were associated with state transitions, a new method was developed. The advanced benefits of the proposed framework were verified by studying the optimal jamming strategies against an 802.11-type wireless network that used the RTS-CTS protocol to communicate and deliver information.

To jam the SU communications without interfering with the PUs, a cognitive jammer with sensing capability can exploit the same statistic information and stochastic dynamic decision-making process that SUs would follow. To this end, an anti-jamming multi-channel access problem was formulated in \cite{jam09} as a non-stochastic multi-armed bandit problem. By taking advantage of shared information among the transceivers, a protocol was developed that enabled SUs to selectively sense channels with a high probability of non-occupancy by jammers and PUs based on the sensing and access historical information.

{To proactively avoid jammed channels, Q-learning was employed to learn strategies of jammers in \cite{jam05} \cite{jam06}.} Due to the time-consuming training process required by Q learning for learning the behaviors of jammers, a wideband spectrum sensing ability was adopted to speed up the learning process. Prior learned information was also used to minimize the number of collisions with the jammer in the training phase. Finally, the effectiveness and improvement of the modified algorithm were verified.

As shown in Fig. \ref{Jam1}, an adversarial ML approach launching jamming attacks and introducing a defense strategy was presented in \cite{jam012} \cite{jam011}. A transmitter $T$ first sensed channels and identified spectrum opportunities, then transmitted data in idle channels. In the meantime, an attacker $A$ also sensed channels and identified busy channels with the intention of jamming legitimate users' transmissions. A pre-trained ML algorithm was implemented at $T$ to classify a channel as idle or busy. This classifier was unknown to the attacker, while $A$ also sensed the channel to capture $T$'s decisions by tracking the acknowledgments. By applying a DL with inference attack, the attacker also built a classifier that was functionally equivalent to the one at the transmitter. Therefore, $A$ could reliably predict successful transmissions based on the sensing results and effectively jam these transmissions. By exploiting the sensitivity of DL to training errors, a defense scheme was then developed by $T$ to defend adversarial DL. The transmitter deliberately used a small number of wrong actions to launch a poisoning attack on the attacker when it accesses the spectrum. The goal is to prevent $A$ from building a reliable classifier. To this end, $T$ systematically decided when to take wrong actions to balance the conflicting effects of deceiving $A$ and making correct transmission decisions. This defense scheme successfully fools the attacker into making prediction errors and allows the transmitter to sustain its performance under intelligent jamming attacks.

\subsubsection{AmBC Empowered Anti-Jamming Methods}
Users in the AmBC network are vulnerable to interference and jamming since their operations are based on ambient RF signals with a limited power supply. However, every cloud has a silver lining. The jamming attacks can be used as additional sources  of energy and information by AmBC empowered systems.

To observe the performance of the AmBC system under a jamming attack, the interaction between a user and an intelligent jammer was modeled as a game in \cite{jam013}. The backscattering time utility functions of both user and jammer were designed, and the closed-form expression for the equilibrium of the Stackelberg game was obtained. As the system SNR  information and transmission strategy of the jammer were not available, Q learning was employed to obtain the optimal strategy in a dynamic iterative manner. Hot booting Q-learning was further introduced to accelerate the convergence of traditional Q learning.

\begin{figure}[h]
	\centering
	\includegraphics[width=3.0in]{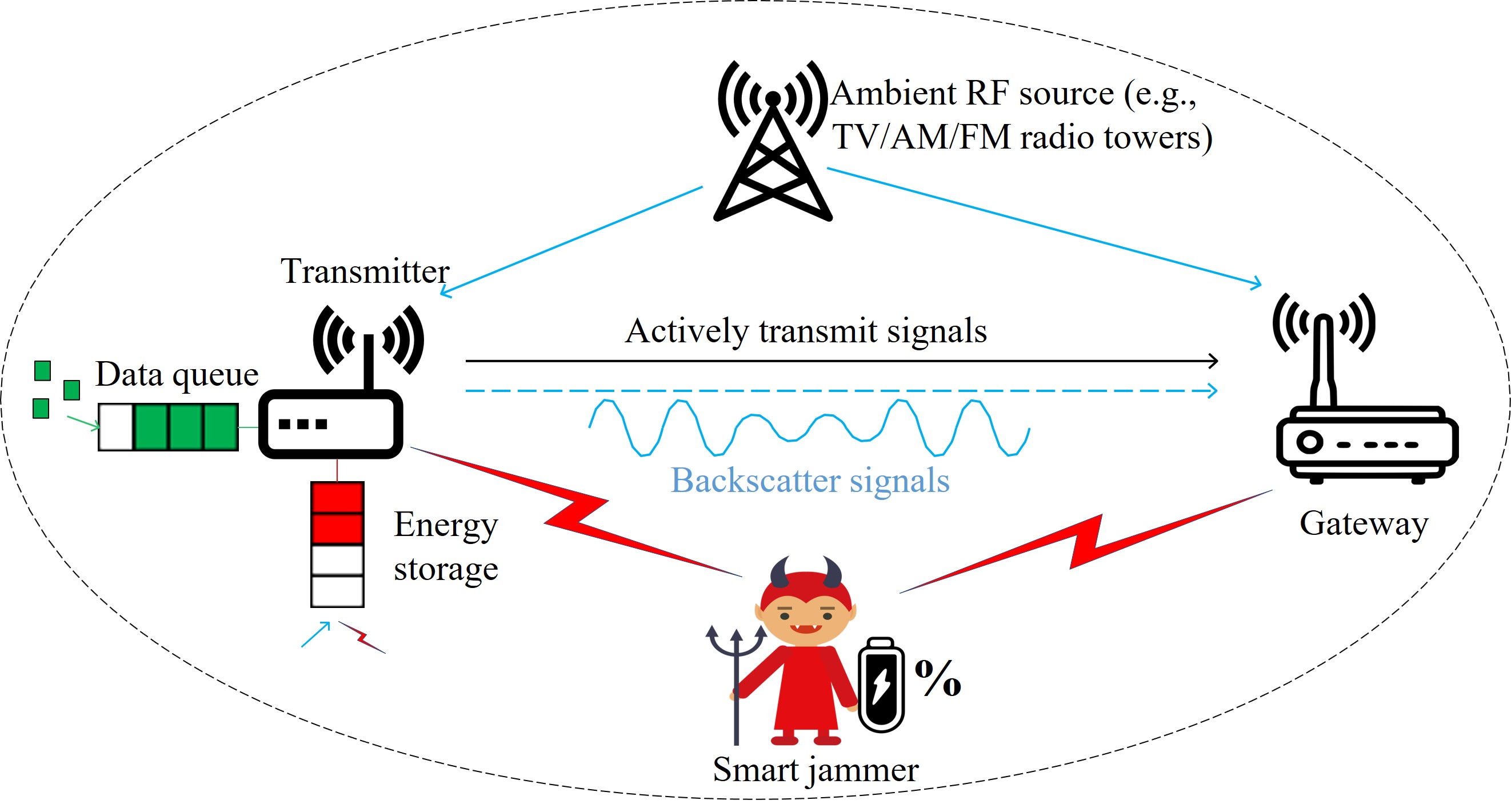}	
	\caption{ {Anti-jamming attack in AmBC-CRN \cite{jam014}.\label{Jam}}}
\end{figure}

Most jamming countermeasures  focus on how to enable users to efficiently escape the invaded channel. AmBC opens the possibility fighting against the malicious jammer. As shown in Fig. \ref{Jam}, a method that allowed wireless nodes to fight against a jamming attack instead of escaping was proposed in \cite{jam014}, \cite{jam016}. By first learning the adversary’s jamming strategy, the users could decide whether or not to adopt the rate or backscatter modulated information on the jamming signals. A dueling neural network architecture-based DRL algorithm was developed to deal with unknown jamming attacks such as jamming strategies, jamming power levels, and jamming capability. The proposed algorithm allowed the transmitter to effectively learn about the jammer and conceive optimal countermeasure actions such as adapting the transmission rate, backscattering, harvesting energy, or staying idle. {The system performance in terms of learning speed, throughput, and packet loss were all significantly improved by the proposed algorithm.}

An RL-based jamming defense method was developed in \cite{jam015}, where the transmitter could obtain the optimal operation policy through real-time interaction processes with the malicious attacker. To be specific, when the jammer attacked the channel, the transmitter could leverage the jamming signals to transmit data by using the ambient backscatter technique or harvest energy from the jamming signals to support its operation. {Thus, the proposed method enabled the transmitter to transmit data even under jamming attacks.} It was also observed that the more power the jammer used to attack the channel, the better the network performed.

\subsection{State-of-the-Art Methods of Defense Against Intercept/Eavesdrop}
In eavesdropping attacks, an attacker tries to intercept private information from the legalized communication system. The basic principle of all defense methods is keeping leakage of information at an acceptable level. The encryption methods aim to totally block this leakage while the physical layer security methods maintain the leakage rate under the required threshold by using different methods such as providing a higher channel difference, introducing friendly jammers, and/or adding artificial noise. Many ML-based works have been proposed to defend against eavesdropping attacks.

\subsubsection{Game Theory Based Anti-Eavesdropping Methods}

When an SS network is invaded by eavesdroppers, similar to jamming defense methods, receivers can always avoid the attack through dynamic access to different spectrum bands. Since there is never cooperation or communication between receivers and eavesdroppers, the attack-avoid process is usually formed as a multi-agent game model. Many game theory-based methods have been proposed.

Secure communication for SUs was studied in \cite{eve02} with the presence of several eavesdroppers and PUs. The interactions between SUs and eavesdroppers were first analyzed through a non-cooperative game framework. {A secure channel selection algorithm was then proposed to enable the SUs and eavesdroppers to make distributed decisions to reach the Nash equilibrium point.} As a result, it was shown that the secrecy rate for each SU was improved by $32.7\%$, and the computation needed to reach Nash equilibrium was reduced by up to $86.5\%$.

The joint physical layer security and EE optimization problems were investigated in \cite{eve03}. The best strategies of power control and relay’s cooperation needed to be determined for both decode-and-forward and amplify-and-forward protocols. By considering the existence of multiple eavesdroppers, a One-Leader One-Follower Stackelberg (OLOFS) game was developed to seek the optimal power allocation and pricing strategy for maximizing the users’ utilities. Moreover, with the consideration of perfect and imperfect CSI information, the best relay selection criteria for the OLOFS game were carried out to maximize the secrecy capacity of the network. Finally, a stochastic learning automata-based distributed learning algorithm was proposed to reach the game's equilibrium, and its benefits to security performance was verified.

Eavesdropper may also target the primary system, and legitimate SUs can act as friendly jammers to alleviate the leakage of PUs' critical information. A SU assisted anti-eavesdropping communication method was investigated in \cite{eve04} for protection of the primary system. A Stackelberg game-based cooperative jamming approach was developed by letting the PUs acted as game leaders and SUs acted as followers. According to a pre-determined probability, SUs send the jamming signals to maximize the data rate priced by the vested power. {To increase the willingness of SUs to provide the friendly jamming service, the system needed to offer more accessible spectrum opportunities for SUs.} The traffic load of PUs needed to be tuned give SUs access to spectrum holes. A continuous-time Markov chain was adopted to model the evolutionary behaviors of the system. The optimal strategies of PUs and SUs as Stackelberg equilibrium were analyzed in the proposed scheme.

To defend against different malicious attacks and build secure networks, a data-driven self-awareness (SA) module was incorporated into the CRN in \cite{eve09}. In this system, the smart attackers could manipulate the radio band to teach the CR wrong behaviors and perform mistaken decisions. At the same time, a basic SA module was found to help the system learn generative models and become aware of abnormal activities inside the radio spectrum. In particular, two real-world practical methods based on different data dimensionality and sampling rates were presented. The first one aimed to investigate the dynamic behaviors of the wideband signal, where the Conditional Generative Adversarial Network (C-GAN) was employed to deal with the high dimensional generalize state vectors extracted from the spectrum representation samples. {The second method adopted a dynamic Bayesian network model to learn the low dimensional generalized state vectors that contained sub-band information.} The effectiveness of proposed approaches for abnormal signal detection was verified using both the real mmWave data set and the simulated OFDM dataset.

\subsubsection{RL Based Anti-Eavesdropping Methods}

A multilevel Stackelberg game based secrecy transmission of CRN under an eavesdropper attack was considered in \cite{eve05}. To protect the achievable rate, some SUs acted as the trusted decode and forward relays. Moreover, to proactively protect the legitimated transceivers, some SUs offered friendly jamming services and requested corresponding service charge prices. Furthermore, an advanced encryption method was adopted to increase the effective security level when users accessed the primary spectrum in the presence of eavesdroppers. By this means, the achievable rate was maximized and the consumed power minimized. Finally, a fuzzy-based MDP Outcome Prediction (MDPOP) Q learning algorithm was proposed to eradicate eavesdropping occurrence in CRNs.

In \cite{eve010}, a DRL-based relay selection for secure buffer-aided CRNs was investigated. Considering that an eavesdropper keeps intercepting the signals from the source and relays, the relay selection problem was modeled as an MDP problem to protect the transmission data. A DQN based approach was introduced to solve this MDP problem, and the $\epsilon$-greedy strategy was applied to balance the exploitation and exploration.

A secure EE based communications problem with energy harvesting ability in CRN was investigated in \cite{eve07}. With the limited energy supply and presence of passive eavesdroppers, a TL actor-critic learning-based algorithm was introduced to help the SUs determine their operation mode to achieve a higher security level. In particular, SUs interact with the environment directly and choose to either stay idle to save energy or transmit the encrypted sensing results to FC  by using a suitable private-key encryption method to maximize the long-term effective security level of the network.

In real applications, some government agencies need to locate suspicious communications via legitimate eavesdropping in an efficient manner. To this end, it is necessary to study the optimal attack strategies for energy-constrained eavesdroppers. A full-duplex active eavesdropper with a limited energy budget was considered in \cite{eve08}. It sought to capture data and interfere with suspicious transmission links. A legitimate attack optimization problem was formulated based on a partially observable MDP framework to maximize the achievable wiretap rate while minimizing the suspicious throughput over a Rayleigh fading channel. Based on the available energy and beliefs regarding licensed channel activity, eavesdroppers needed to determine the course of action with maximum long-term system benefits. This may be either passive eavesdropping without jamming or active eavesdropping with an optimal amount of jamming energy.

\subsubsection{AmBC based Methods}
 Because of their simple coding and modulation schemes, backscatter communications are quite vulnerable to security attacks such as eavesdropping. The passive nature of backscatter communications make it challenging to protect their secrecy. Moreover, a special challenge in AmBC arises from its limited computational and communication resources, which dictate that the authentication and encryption methods must be lightweight. To meet these constraints, physical layer security methods can be applied to protect the security of AmBC links \cite{amb02}.

\begin{figure}[h]
	\centering
	\includegraphics[width=3.0in]{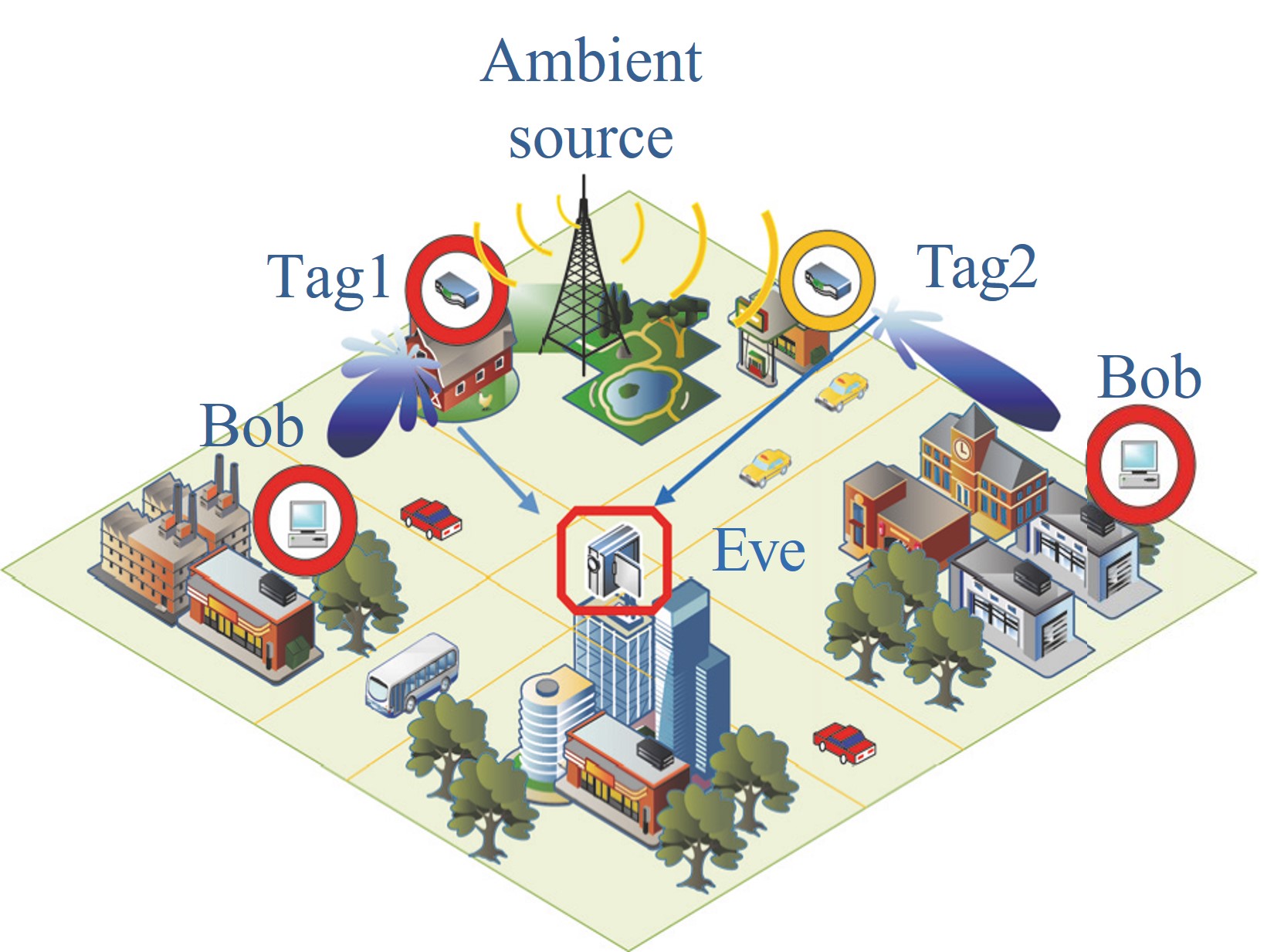}	
	\caption{ {Influence of antenna side lobes on communication. \cite{eve01}.\label{eve1}}}
\end{figure}

As shown in Fig. \ref{eve1},  nondirectional forms of communication in AmBC networks are prone to information leakage. Reducing the side lobe level is therefore crucial to preventing eavesdropping. {To this end, an ML-based antenna design scheme was proposed in \cite{eve01} that achieved directional communication between transceivers by combining patch antenna with Log Periodic Dual-dipole Antenna (LPDA). Aiming to limit the number of large side lobes and reduce the Side Lobe Level (SLL), a multi-objective genetic algorithm was proposed to optimize the antenna side lobe, gain, standing wave ratio, and return loss.} It was shown the proposed method could significantly reduce information leakage while guaranteeing communication quality.

\subsection{State-of-the-Art ML Based Privacy Protection Methods}
In this subsection, privacy protections for PU, SU, and ML in different SS frameworks are sequentially reviewed.

\subsubsection{Privacy Protection for PUs in SS Networks}
In some SS networks such as SAS and LSA, leakage of PU's privacy can cause serious security threats. Potential malicious adversaries may exploit attacking technologies such as DIA to obtain information about the IU and that can later be used to develop attack strategies.  
To defend the DIA and protect IUs, one viable approach is to obfuscate the information revealed by the database. There are a lot of works used ML-based obfuscation techniques to counter inference attacks.

The authors in \cite{prss03} investigated whether or not a malicious opponent could infer the radar's location through veiled radar information contained in the system's precoder. An ML-based location inference attack was simulated for two specific precoder schemes. It was verified that radar privacy could be compromised by system information leaked through the precoder, introducing various degrees of risks.

The tradeoff between privacy preservation and spectrum efficiency was examined in \cite{prss04}. A generalized SS system architecture was proposed and a multi-utility user privacy optimization problem was formulated. Potential adversary inference attacks were used to measure privacy, and an efficient heuristic strategy was developed to solve the problem. Compared with existing obfuscation strategies, the proposed approach can achieved a $50\%$ increase in privacy with an insignificant impact on SE.

In \cite{dbss1}, the authors first showed that adversarial SUs could employ a Bayesian learning-based inference algorithm to accurately locate a non-stationary radar system using only information gathered from seemingly innocuous query replies obtained from a SAS.  Several obfuscation techniques were then proposed and implemented in the SAS for countering such inference attacks. Finally, the obfuscation techniques' efficacy in minimizing spectral efficiency loss while preserving incumbent privacy was investigated.

The authors in \cite{gbss03} proposed a CS-based federated learning framework to achieve IU detection for improving communication efficiency while protecting the privacy of training samples. By using an MMV CS model, each sensor transmitted the updated aggregated parameters instead of the raw spectrum data to the central server to protect privacy. {They demonstrated that the detection performance was as good as the scheme under the raw training samples, while significantly improving  the communication and training efficiency.}
  
\subsubsection{Privacy Protection for SUs in SS Networks}
{
In \cite{prss01}, several location privacy-related attacks in CSS-based CRNs were first identified. Such attacks can threaten SUs' location privacy by correlating their sensing reports and their physical location. To prevent leakage of location privacy, a privacy-preserving framework was proposed. It was demonstrated that without efficient protection, the attackers could compromise a SU’s location privacy at a success rate of more than $90\%$. 
{A proposed privacy-preserving framework was further introduced and verified, which could significantly improve the location privacy of SUs with a minimal effect on the performance of collaborative sensing.}

To protect the location privacy of SUs while allowing them to sense vicinity spectrum availability, two location privacy-preserving schemes for database-driven CRNs were studied in \cite{gbss04}. 
The spectrum databases' structured nature and SUs' queries were exploited by those schemes to create a compact representation of databases that could be queried by SUs without requiring them to share their location with the database, thereby eliminating the possibility of location leakage. Based on whether a user is a member of a set or not, the first method, location privacy in database-driven CRNs (LPDB), constructed a compact version of the database and provided optimal location privacy to SUs in the coverage area. It achieved unconditional security with an acceptable communication overhead. The second method, LPDB with two servers (LPDBQS), minimized SUs' overhead with an additional network entity cost. The tradeoff between cost and performance provided more options for system design based on specific requirements.

In \cite{prss02}, an aggregative game was used to model SS in large-scale, heterogeneous, and dynamic network. By utilizing past channel access experience, an online learning algorithm was proposed to improve the utility of each user. Considering the heterogeneous impact of users, a multi-dimensional aggregative game was used to model the SS of the large-scale wireless network. A mediated privacy-preserving and truthful mechanism were developed to achieve an $\eta$-approximate ex-post NE and provided no regret guarantee for each user. It was shown that the proposed method satisfied joint differential privacy. 
 }

 
\subsubsection{Privacy Protection for ML Algorithms}

The sensitive training data in ML-based applications faced distinct privacy issues. Malicious attackers can obtain private information through the structure of models or their observations. To investigate the privacy leakage of training data, the authors in \cite{mlpr01} introduced novel formal definitions of advantage for membership and attribute inference attacks. {Attacks in different learning algorithms and model properties were analyzed based on these definitions.} It was shown that overfit could increase the risk of privacy leakages.

{Secure Multi Party Computation (MPC) allows different entities to share joint data to train their models without releasing any private information in the training data.} The MPC-based privacy-preserving method was investigated in \cite{mlss02} for linear regression, logistic regression, and neural network training using the stochastic gradient descent method. A two-server model was considered, and the training data was securely distributed among two non-colluding servers. Different models were trained on the joint data using secure Two Party Computation (2PC). It was shown that their new techniques could significantly increase speed while guaranteeing performance without leaking data privacy.

Cloud computing frameworks provide many benefits to the communication networks, such as powerful processability and unlimited storage space. {However, some cloud services are provided by third parties such as Amazon AWS, Google Cloud, Microsoft Azure, etc. 
Users may hesitate to entrust their sensitive data to these entities, and rightly so.}
To this end, equipping cloud computing-assisted ML with effective protection methods is critical.

To protect different data owners' privacy, the authors in \cite{clpr00} introduced a new efficient method that allowed all participants to publicly verify the veracity of the encrypted data. A Unidirectional Proxy Re-Encryption (UPRE) method was also adopted to lower the computation costs. A noise was further added to the encrypted data to preserve the private information while guaranteeing the effectiveness of ML training on cloud.

A cloud-assisted privacy-preserving ML framework was developed in \cite{clpr01}. By using outsourced ML algorithms, the cloud server first generated a model, then processed testing data from the network with the generated model in real-time. The proposed framework adopted a differential privacy method of performing privacy-preserving data analysis and homomorphic encryption in order to conduct valid operations over encrypted data.

FL allowed decoupling of data provision and ML model aggregation and shows promise as a framework for addressing  privacy problems for distributed ML \cite{hjxiang}. 
It enables the users to cooperatively learn a global model without sacrificing data privacy directly.  The information transmitted for FL consists of minimal updates to improve a particular machine learning model. However, the design of FL still needs the protection of parameters as well as investigations on the tradeoffs between the privacy-security-level and the system performance. The study \cite{flpr00} suggested that FL could expose intermediate results such as stochastic gradient descent, and the transmission of these gradients may actually leak private information when exposed together with a data structure. It is still possible for adversaries to reconstruct the raw data approximately, especially when the architecture and parameters are not completely protected.

To investigate the leakage of private information in users’ data, the performance of malicious servers was studied in  \cite{fl02}. A GAN-based framework with a multi-task discriminator capable of discriminating category, reality, and client identity included in input samples simultaneously was introduced. It was shown that the generator could easily recover the specific private data of users , particularly client identity.

An efficient and robust protocol for high-dimensional data secure aggregation was proposed in \cite{flpr01} that can be used in FL. Using this protocol, a server was able to compute the sum of large user-held data vectors from mobile devices to aggregate user-provided model updates for a DNN model without distinguishing individual user's contributions. In addition, the effectiveness and efficiency of the proposed protocol were verified.

A GAN-based privacy-preserving method was proposed in \cite{dbss2} to obfuscate users' sensitive information. The proposed method employed a generator to produce an optimal obfuscation method for data protection. At the same time, a classifier was used to deobfuscate the data. These two nets continued to play against each other until they achieved an equilibrium. This process can raise the level of protection. By investigating location privacy protection on the Gowalla dataset and synthetic data, it was shown that the proposed approach could achieve privacy protection and deal with the Bayes error.

To alleviate the threat of black-box inference attacks against ML models, a mechanism to train models with membership privacy was introduced in \cite{mlpr003}. By formulating a min-max game, an adversarial training algorithm was designed to minimize the prediction loss of the model and the maximum gain of the inference attacks. The effectiveness of the min-max strategy on defending membership inference attacks was verified without significantly downgrading the model's prediction accuracy.

To defend against the attribute inference attacks, a countermeasure named AttriGuard was proposed in \cite{dbss3}. AttriGuard works in two phases. In Phase I, the minimum noise was found by adapting existing evasion attacks in adversarial ML. This noise protects users' attribute values by adding itself to the user’s public data.  
In Phase II, the proposed method sampled one attribute value according to a certain probability distribution and added the corresponding noise found in Phase I to the user’s public data.


\section{Open Issues and Future Challenges}
In this section, open issues related to ML-based SS techniques are discussed.
\subsection{Cross Technology Coexistence}

Cross-technology coexistence design is an important research direction for future SS and security protection.
Although we have discussed different SS frameworks separately, cross-technology coexistence design has not been fully investigated. Problems associated with the combination and coordination of different technologies provide ample material for future studies. As shown in \cite{amb13, amb14, jam012, jam011}, combining CRN and AmBC can provide the system with multiple sharing capabilities and the possibility of countering attacks. Exploring the combination of different SS technologies could help design a more efficient sharing system. 
{
Studying the characteristics of different ML solutions and designing a combination SS solution more in line with the characteristics and requirements of the system could further improve performance.
For example, the ensemble learning used in the literature \cite{ssdf24,ssdf25} improved the efficiency of the system by combining linear regression, neural networks, and other solutions, allowing it to combat PUE and SSDF quickly and efficiently.} There is a real need for research on new protocols and frameworks for such coexistence.

\subsection{Advanced Signal Processing Techniques}
SS inevitably further complicates the already challenging radio environment. The traditional signal process techniques designed for exclusive spectrum applications may not suit the new SS system. Advanced signal processing can help quickly extract discriminative statistical features of all emitters and facilitate the system to implement new SS techniques.  ML also allows the system to efficiently pick up and classify the signals based on their features. Moreover, GAN and other learning algorithms offer the possibility for the system to generate signal information, allowing greater potential for the signal processing design \cite{cr1202 , eve09, fl02, puedc13}. Therefore, new signal processing methods are desirable to make the SS system a viable solution for future generation communications.


\subsection{Energy Efficiency}
One critical issue in the SS network is EE. On one hand, densely deployed devices in 5G quickly increase spectrum scarcity and energy consumption. On the other hand, the limited battery capacity of mobile devices and IoT devices necessitates an energy-efficient solution to support the massive data communication requirement and maintain long-term performance \cite{qunee}. Moreover, as the devices in the SS system need additional operations, such as spectrum sensing, to maintain dynamic access than the exclusively used spectrum network, the energy consumption is higher than that of traditional accessing paradigms. Therefore, new EE-based schemes need to be explored for the complex ML-based SS scenarios.


\subsection{Fairness} 
Since the SS networks involve competition for the same spectrum resources, how to guarantee fairness among users is critical to providing a better QoS. {For some systems such as LTE-U/LTE-LAA, fairness between different systems has been considered by using specific mechanisms like the duty cycle \cite{lteu04, lteu05, lteu010}.} However, fairness among SUs remains an open issue for other SS frameworks and requires further research. Different operators may also put the original exclusive spectrum into a spectrum pool for sharing in next-generation communications, eliminating the priority of the ownership of the spectrum. The question of how to use ML to achieve a fair sharing solution needs more study.

 {
\subsection{Explainable ML in SS networks}
The combination of SS networks with ML algorithms has shown a lot of advantages compared to the classical optimization theories. However, in most cases, an ML algorithm is treated as a black box, which raises concerns on potential risks when designing a practical system with sensitive applications. Therefore, it is demanded to understand the specific model operation and the principle behind this model, especially for the optimization decisions in security-related issues in the SS network. Therefore, explainability is a prerequisite to ensure the desired outcome.
}

 {
The research of explainable artificial intelligence (AI), informed ML, or intelligible intelligence in SS networks and the related defense solutions, is therefore very important and remains an open issue that needs extensive future work \cite{explml}.
}
\subsection{Upper Layer Security Issues}
Most of the security threats and countermeasures covered in this paper focused on the physical and MAC layers. Security concerns for upper-layer protocol design in SS call for further investigations as well.
Network layer attacks like Hello flood, acknowledgment spoofing, application layer attacks such as software viruses, and malware transfer of malicious code all need to be investigated in the context of SS systems.
Furthermore, the cross-layer security design involving the interaction between the physical layer and upper layers also presents many open issues. 

\subsection{Intelligent Attackers} 
Many existing works assume that the attackers still employ traditional attack strategies. However, as virtue rises one foot, vice rises ten. Advanced ML algorithms can also allow attackers to upgrade their attack strategies\cite{jam04, jam09, jam05, jam06, jam012, jam011}. ML can be exploited to find the best strategies for blocking spectrum access or transmission. For example, DL can be used to predict the activity of users. DRL can be used to choose the best channel and attack strategies based on users’ behaviors. Defense measures using traditional attack assumptions may become ineffective against intelligent attackers. Understanding the optimal attacking strategies helps quantify the severity or impact of an attacker on the system and sheds light on the design of defence strategies.

\subsection{Composite Security Attacks}
Attackers will try to disturb and break the system by using any possible methods. {For instance, attackers may first try to eavesdrop on information from legal users. If blocked by a friendly jammer, they may switch from a passive intercept mode to an active attack mode, sending jamming signals to interfere with information transmission between legitimate transmitters and receivers.}
Since the defence is always passively resisting, how to effectively fight against mixed attack strategies and quickly respond to changes in attack mode is an interesting problem that must be studied. Other possible problems arise when considering the possibility of dealing with multiple attacks once.
Two separate attackers simultaneously delivering, say, a jamming and an eavesdropping attack, might cancel each other’s jamming signals.
These composite security attack issues require future study. 

\subsection{Attacks for ML model}
\begin{figure}[h]
	\centering
	\includegraphics[width=3in]{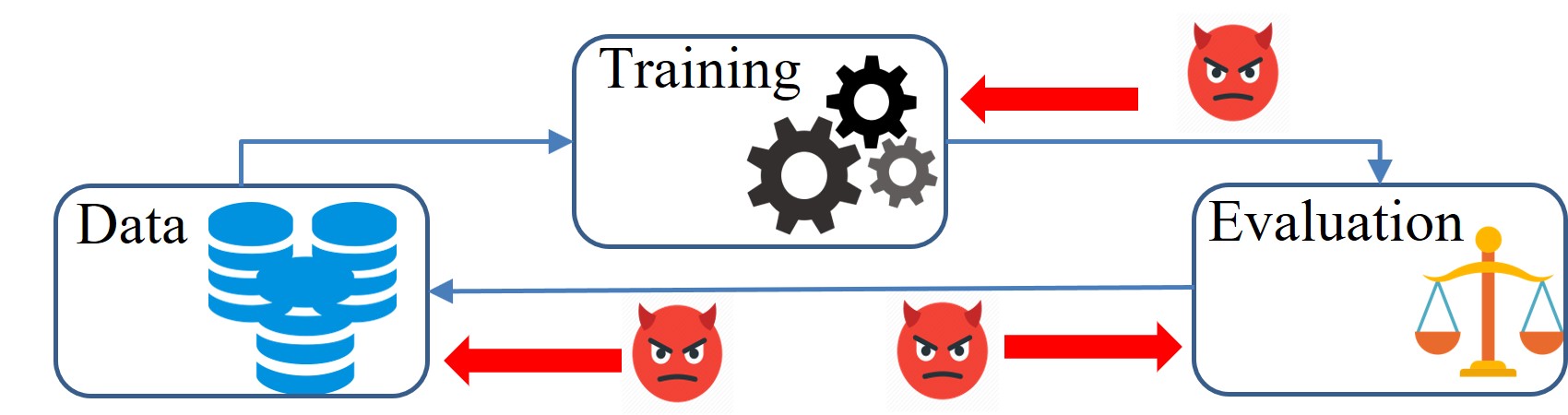}
	\caption{ {Illustration of attacks to ML model.\label{MLattack}}}
\end{figure}

Besides launching attacks based on the SS framework, attackers can also take advantage of opportunities in ML workflow to disturb the system operation. 
 {
As shown in Fig. \ref{MLattack}, a typical ML using data to train a model and then evaluating the trained model with the test data can expose its workflow to various types of attacks: Exploratory attacks, Evasion attacks, Poisoning attacks, Backdoor attacks, etc\cite{ssdf19}.
}

 {
Exploratory attacks, also called inference attacks,  discover how the underlying ML works for an application. It usually maintains a surrogate model to mock the victim ML system with the same input and output data types. Exploratory attacks aim to infer sensitive and proprietary information of victim systems to launch vast subsequent attacks to it. There are limited existing works that studied exploratory attacks for CRN networks and corresponding defense methods \cite{jam012}. The attackers can sense the victims' activities to build an ML model, and defenders can also deliberately mislead the attackers' model.
}

 {  
Evasion attackers might trick the ML algorithm into making wrong decisions, such as fooling a security algorithm into accepting an adversary as legitimate. It can be achieved by manipulating the test data to mislead the model. Existing research on evasion attacks  in SS communication systems mainly focuses on misleading the classifier at the receiver side by launching the spectrum poisoning attacks during the sensing phase. It aims to change the channel status features and forces the system to make wrong transmission decisions. It should be noted that this attack differs from SSDF attacks because the attackers mainly focus on injecting adversarial perturbations over the air to the channel instead participated in CSS \cite{ssdf19} \cite{even1}.
}

 {  
Poisoning attacks, also called causative attacks, provide incorrect information such as training data to the ML to cause the ML model to perform poorly. The poisoning attacks in the context of SS networks have been investigated in \cite{pos}, which can be achieved by fooling the classifiers with spectrum data falsifications during the CSS phase, similar to SSDF attacks.
}

 {
Backdoor attackers train the ML model by deliberately misclassifying any input with an added trigger to a specific target label. The attackers need to first construct the backdoored data that contains a trigger within a subset of clean data and change their labels to the target one. They then mix this backdoored data with clean data to train the model to learn the original tasks and backdoor behaviors \cite{backdoor}. Backdoor attacks can be exploited to help the attackers to pass the authentication system and grant unauthorized access right. This can cause severe security and privacy consequences for ML-based SS networks, such as  ML-based database-assisted SS systems and distributed ML models-based defense approaches.
}

 {
Researchers need to work on possible attack threats and defense strategies based on SS communication scenarios.
}
\section{Conclusions}
 {
In this paper, to obtain insight into the combination of ML algorithms and SS communications, four SS application scenarios were first investigated, i.e., opportunistic access-based CRNs, database-assisted SS systems, LTE-U/LTE-LAA networks, and symbiotic SS mechanism-based AmBC networks. 
How to use ML in tackling SS-related problems was presented.
A comprehensive investigation of state-of-the-art ML-based SS solutions and their performance gains were discussed.
However, it has been noted that the dynamic access and sharing paradigms of SS networks may open the system to many security concerns. Correspondingly two typical spectrum sensing attacks were discussed, i.e., PUE and SSDF. Two common attacks,  i.e., jamming and eavesdropping, during wireless access and transmission in the context of the SS network were also addressed. Furthermore, connecting a large number of users and the application of ML all require  massive information exchanges, generating tremendous concerns about privacy. The paper further presented the state-of-art research on privacy protection for SUs and PUs, as well as the use of ML mechanisms for that purpose.  Finally, the open issues and new challenges in SS networks were highlighted for future research.
}
\\
\bibliographystyle{IEEEtran}
\bibliography{lib}

\end{document}